\documentclass[%
 reprint,
superscriptaddress,
 amsmath,amssymb,
 aps,
pra,
]{revtex4-2}
\usepackage{CJKutf8}
\usepackage{lineno,hyperref}
\usepackage[squaren]{SIunits}
\usepackage{color}
\usepackage{amsmath}
\usepackage{caption,subcaption}
\usepackage{setspace}
\usepackage{multirow}
\usepackage{threeparttable}
\usepackage{booktabs}
\usepackage{graphicx}% Include figure files
\usepackage{dcolumn}% Align table columns on decimal point
\usepackage{bm}% bold math
\definecolor{americanrose}{rgb}{1.0, 0.01, 0.24}
\definecolor{coralpink}{rgb}{0.97, 0.51, 0.47}
\definecolor{ao(english)}{rgb}{0.0, 0.5, 0.0}
\definecolor{darkpastelgreen}{rgb}{0.01, 0.75, 0.24}
\definecolor{cyan(process)}{rgb}{0.0, 0.72, 0.92}

\newcommand\isoindoleHpyrrole{S15}

\begin{document}

\preprint{APS/123-QED}

\title{Vibronic fine structure in the nitrogen 1s photoelectron spectra from Franck-Condon simulations. III.  Rules for amine/imine N atoms in small N-heterocycles
}

 \author{Minrui Wei}
 \affiliation{MIIT Key Laboratory of Semiconductor Microstructure and Quantum Sensing, Department of Applied Physics, School of Physics, Nanjing University of Science and Technology, 210094 Nanjing, China}
 
 \author{Junxiang Zuo}
 \affiliation{MIIT Key Laboratory of Semiconductor Microstructure and Quantum Sensing, Department of Applied Physics, School of Physics, Nanjing University of Science and Technology, 210094 Nanjing, China}

 \author{Guangjun Tian}%
 \email{tian@ysu.edu.cn}
\affiliation{%
 Key Laboratory for Microstructural Material Physics of Hebei Province, School of Science, Yanshan University, 066004 Qinhuangdao, China}%
 
 \author{Weijie Hua}%
 \email{wjhua@njust.edu.cn}
 \affiliation{MIIT Key Laboratory of Semiconductor Microstructure and Quantum Sensing, Department of Applied Physics, School of Physics, Nanjing University of Science and Technology, 210094 Nanjing, China}

\date{\today}% It is always \today, today,
%\begin{CJK*}{UTF8}{gbsn}
\begin{abstract}
Vibronic coupling plays a crucial role in X-ray photoelectron spectra (XPS) of molecules. In a series of three papers, we present a comprehensive exploration of the N-heterocycles family, known for their diverse structures, to summarize the general rules of vibronic coupling in high-resolution vibrationally-resolved XPS spectra at the N1s edge. Building upon our previous studies on six-membered monocyclic azines [Phys. Rev. A 106, 022811 (2022)] and fused bicyclic compounds indoles with five and six members [Phys. Rev. A 108, 022816 (2023)], in this study, we focus on investigating a series of 12 five-membered N-heterocycles using Franck-Condon simulations, incorporating Duschinsky rotation effects and density functional theory. Our calculations reveal distinct spectral characteristics of amine and imine within these 12 systems in binding energies, spectral characteristics, structural changes, vibrational coupling strengths, and effects of hydrogenation. Furthermore, we expand our analysis to encompass all 35 N-heterocycles discussed in the three papers and consolidate these findings into the general rules. we find that 1s ionization in amine nitrogen induces more substantial geometrical changes, resulting in larger vibronic coupling strength compared to imine nitrogens. The spectra of imine nitrogens exhibit two distinct characteristic peaks originating from the 0-0 and 0-1 transitions, whereas the spectra of amine nitrogens are characterized by a broad peak with numerous weak fingerprints due to significant mixing of various 0-$n$ transitions. We observe that amine (imine) nitrogens generally cause a negative (positive) change in zero-point vibrational energy. This study provides valuable insights into vibronic coupling in N-heterocycles, shedding light on the distinguishing features and behavior of amine and imine nitrogens in vibrationally-resolved XPS spectra. 

  \end{abstract}
  
\maketitle

%---------------------------------------------------------------------------------
\section{Introduction} 
%---------------------------------------------------------------------------------

X-ray photoelectron spectroscopy (XPS) is one of the most successful spectroscopic techniques and is widely used in molecules and materials. Existing experimental databases for XPS\cite{Rumble1992NIST, jolly_core-electron_1984, wagner_1979_handbook, interxps, XPSlibrary, espectra, sasj, LaSurface} collect important reference data for a large number of molecules and materials, which facilitate spectral interpretation and help to interpret the physical and chemical insights behind the data.\cite{siegbahn_electron_1982, watts2003introduction, sokolowski1957magnetic} However, most libraries report only inconsistent binding energy (BE) data and important profile information is missing. A high-resolution theoretical library is urgently needed to reveal the physical nature and provide a better reference for the structural characterization.

Vibronic coupling plays an important role in XPS which defines its high-resolution fine structures. To understand the physical nature of vibronic coupling in molecular XPS spectra, we focused on the nitrogen-heterocycle compounds and performed accurate theoretical calculations for a number of molecules. These compounds, which only contain three elements, C, N, and H, are important building blocks in the fields of photochemistry, biology, materials science, and pharmacy, and exhibit rich structural flexibility to study structure-spectroscopy relationships. The search for the rules of vibronic coupling is a necessary preliminary study for the automatic construction of a theoretical library with a large number of molecules. Because the computational method for excited states with a core hole is challenging and sometimes even suffers from convergence problems in the calculation of self-consistent field and geometric optimization. 

We would like to follow up on the early pioneering work of Kai Siegbahn,\cite{siegbahn1982electron, siegbahn1967atomic, siegbahn1974esca, allison_molecular_1972, gelius_molecular_1970, fellner-feldegg_esca_1975, siegbarn_esca_1974} who investigated families of systems by high-resolution experiments and calculations, and simulate the accurate vibrationally-resolved XPS of molecules with similar structures on the same basis. In previous papers of this series, we have systematically studied azines\cite{wei_vibronic_2022} (paper I) and indoles \cite{wei_vibronic_2023} (paper II). These are monocyclic compounds with six-membered rings and bicyclic compounds consisting of fused five- and six-membered rings, respectively. Good agreement was obtained with the experiments, and some general structure-spectroscopy rules were established. Therefore, it is natural and necessary to investigate the performance of pure five-membered ring compounds in this work (paper III).

A total of 12 monocyclic compounds with five-membered rings were selected in this study. These include two pyrrole and three pyrroline isomers, pyrazoline, two diazoles (imidazole and pyrazole), three triazoles (1,3,4-triazole, 1,2,5-triazole, and 1,2,4-triazole), and one anion \textit{cyclo}-N$_5^-$. Pyrrole and imidazole are essential building block molecules in biological systems and are found in compounds, such as amino acids (e.g., tryptophan and histidine) and porphyrins.\cite{alongi_new_2005, wurtz_fmoc_2001, crossley_fused_1999} Pyrroline, on the other hand, is commonly found in natural products, drug molecules, and synthetic intermediates. It has various applications in medicine\cite{gomez_phosphine_2013, nguyen_synthesis_2022, sun_sulfur-directed_2020, anderson_3-pyrroline_1987} and pesticides.\cite{chen_novel_2020, anderson_3-pyrroline_1987} Although pyrazole is rare in nature, its derivatives exhibit diverse pharmacological activities.\cite{kost_progress_1966} Triazoles have significant potential for applications in nitrogen-rich high-energy density materials (HEDMs) due to their high thermal stability and positive enthalpy of formation.\cite{dippold_study_2013, ma_amino-nitramino_2018} The cyclo-pentazole anion \textit{cyclo}-N$_5^-$ represents an important structural motif and has attracted significant research interest in recent years.\cite{wang_recent_2018, zhang_synthesis_2017, yao_recent_2021}

To our knowledge, XPS experiments\cite{jolly_core-electron_1984} on these compounds are relatively outdated and have low resolution.  Core binding energies were reported for part of systems like pyrrole\cite{jolly_core-electron_1984, gelius_esca_1971, pan_chemisorption_1986, cavell1977site} and imidazole.\cite{nolting_pseudoequivalent_2008} While gas-phase N1s XPS spectra have been measured for pyrrole, the limited precision hampers further vibronic analysis.\cite{jolly_core-electron_1984} In light of this, our study aims to provide accurate and vibrationally-resolved N1s XPS spectra for 12 commonly encountered five-membered heterocyclic complexes. Moreover, we endeavor to establish general rules for small N-heterocycles based on comprehensive data acquired throughout the three papers in this series. Our analyses will encompass chemical shifts, fine structures, vibronic transitions, and active vibrational modes, and will particularly focus on distinguishing features between amine and imine nitrogens.

The structure of this work is organized as follows: Section \ref{sec:method} briefly outlines the computational details employed in this study. Subsequently, in Section \ref{sec:results}, we present the results obtained for the 12 five-membered molecules. In Section \ref{sec:discussion}, we summarize the general vibronic rules derived from the complete dataset encompassing 35 molecules across all three papers. Finally, we provide concluding remarks and an outlook for future research.

%---------------------------------------------------------------------------------
 \section{Computational details} \label{sec:method}
%--------------------------------------------------------------------------------

The computational methods were described in detail previously,\cite{wei_vibronic_2022, wei_vibronic_2023} so that only brief expositions are given here. All electronic structure calculations were performed using the GAMESS-US  package\cite{schmidt_general_1993, *gordon_advances_2005} at the density functional theory (DFT) level with the B3LYP functional\cite{becke_density-functional_1988, *becke_new_1993, *lee_development_1988}.  The vibronic fine structure was calculated using a modified DynaVib package.\cite{DynaVib, hua_theoretical_2020}  Within the harmonic oscillator approximation, the Duschinsky rotation (DR) effect\cite{duschinsky_1937} was included to calculate the Franck-Condon (FC) factors.  The normal modes of the ground ($\mathbf{q}^\prime$) and core-ionized ($\mathbf{q}$) electronic states, described by column vectors, are linearly related by the Duschinsky transformation, $\mathbf{q}^\prime=\mathbf{J}\mathbf{q}+\mathbf{k}$. Here $\mathbf{J}$ and $\mathbf{k}$ refer to the Duschinsky rotation matrix and the displacement vector, respectively. The FC amplitude $\langle 0|0\rangle$ was computed based on vibrational frequencies and displacements, and then amplitudes $\langle 0|n\rangle$ in a recursive manner starting from $\langle 0|0\rangle$.\cite{sharp_franckcondon_1964, ruhoff_recursion_1994, ruhoff_algorithms_2000} The FC factor is computed as the square of the FC amplitude.

From the electronic structure calculations, we obtained the vertical ($I^\text{vert}$) and adiabatic ($I^\text{ad}$) ionic potentials (IPs), the difference of zero-point vibrational energies (ZPE) between the final full core hole (FCH) state and the initial ground state (GS) ($\Delta\varepsilon_0 $), and the  0-0 vibrational transition energy ($E_\text{00}^\text{DR}$):\cite{wei_vibronic_2023, wei_vibronic_2022}
\begin{eqnarray}
I^{\rm{vert}} &=& \it E_{\rm{FCH}}|_{\mathbf{min\ GS}} - E_{\rm{GS}}|_{\mathbf{min\ GS}} + \delta_{\rm{rel}},\\ \label{eq:Iv}
I^{\rm{ad}} &=& \it E_{\rm{FCH}}|_{\mathbf{min\ FCH}} - E_{\rm{GS}}|_{\mathbf{min\ GS}} + \delta_{\rm{rel}},\\ \label{eq:Ia}
\Delta\varepsilon_0 &=& \varepsilon_0^{\rm{FCH}}- \varepsilon_0^{\rm{GS}},\label{eq:deps0}\\
E_\text{00}^\text{DR}&=&  I^{\rm{ad}} + \Delta\varepsilon_0.\label{eq:E00}
\end{eqnarray}
Here $E_{\rm{GS}}$ and $E_{\rm{FCH}}$ stand for the total energies of the GS and  FCH states, respectively. $\mathbf{min\ GS}$ and $\mathbf{min\ FCH}$ denote optimized structures of the two states. To account for the scalar relativistic effect of the N1s core hole, the calculated IPs were calibrated by adding a uniform shift of $\delta_{\rm{rel}}$=0.3 eV.\cite{triguero_separate_1999} Stick spectra were convoluted with a Lorentzian line shape function, where half-width-at-half-maximum (hwhm) values of $\gamma=0.05$ eV were used for most systems. To better visualize the profile details, slightly smaller $\gamma$=0.04 eV for 2-pyrroline and $\gamma$=0.03 eV for pyrrole and {cyclo}-N$_5^-$ were used.  For molecules with multiple nitrogens, the sum of individual atom-specific spectra leads to the total spectrum.
%---------------------------------------------------------------------------------
\section{\label{sec:results}Results for Five-membered N-heterocycles}
%---------------------------------------------------------------------------------

%+++++++++++++++++++++++++++++++
\subsection{Statistics on IPs} 
%+++++++++++++++++++++++++++++++

Table \ref{tab:ip} presents the vertical (adiabatic) N1s IPs calculated by B3LYP for all systems and the visualization is also provided in Fig. S1. The only anion \textit{cyclo}-N$_5^-$ exhibits the lowest vertical (400.6 eV) and adiabatic (400.3 eV) IPs, which differs significantly from the remaining molecules. For these molecules, there are a total of 20 N centers, comprising 9 amine and 11 imine nitrogen atoms. The vertical (adiabatic) IPs range from 404.2 to 407.9 eV (404.1 to 407.4 eV) spanning a range of 3.7 eV (3.1 eV).  Amine and imine nitrogens can be readily distinguished based on their IPs: amine N atoms generally show larger vertical (adiabatic) IPs of 404.5--407.9 eV (404.2--407.0 eV), whereas imine N atoms have IPs ranging over 404.2--406.0 eV (404.1--405.8 eV). It should be noted that there is some overlap between the two regions.

The core ionization-induced structural relaxation effect in the excited-state potential energy surface (PES) can be estimated by calculating the difference between vertical and adiabatic IPs, defined as:
\begin{eqnarray} 
\Delta I &\equiv& I^\text{vert} - I^\text{ad} \nonumber\\ 
&=& E_\text{FCH}|_\mathbf{min\,GS} - E_\text{FCH}|_\mathbf{min\,FCH}.\label{eq:dI}
\end{eqnarray}
This difference of ionization potential, $\Delta I $, is always non-negative. Our calculations revealed that amine nitrogen atoms (0.3--0.5 eV) exhibit larger $\Delta I$ values compared to imine nitrogen atoms (0.1--0.2 eV). This indicates that amine nitrogens undergo larger changes in geometrical structure compared to imine nitrogens.

Our theoretical results agree well with the experiments.\cite{jolly_core-electron_1984, gelius_esca_1971, pan_chemisorption_1986, cavell1977site, nolting_pseudoequivalent_2008}. The deviations ranged from -0.2 to 1.0 eV for the vertical IPs and from -0.4 to 0.6 eV for the adiabatic IPs. These discrepancies are consistent with the typical accuracy range of 0.5--1.0 eV observed in the $\Delta$Kohn-Sham method for predicting 1s binding energies of light elements\cite{bagus_consequences_2016, pueyo_bellafont_validation_2015, pueyo_bellafont_prediction_2015, pueyo_SCF_performance_2016, du_theoretical_2022}.

%E00和zpe
%+++++++++++++++++++++++++++++++
\subsection{Changes in structure}\label{sec:rmsd}
%+++++++++++++++++++++++++++++++

Table \ref{tab:st} summarizes the local structural parameters near the ionized nitrogen (N$^*$) for all molecules at the optimized FCH state and the changes compared to the corresponding optimized ground state. We note that at each N site, the two N--$X$ ($X=\text{C}, \text{N}$) distances (in the ground state) exhibit a more pronounced accordance  with the Kekul\'{e} structures, demonstrating a distinct pattern of one long and one short distance. Let us use N$_\text{a}$ and N$_\text{i}$ to denote an amine and imine nitrogen, respectively. The N$^*$-$X$ bond lengths exhibit distinct difference for the two types: N$_\text{a}^*$-$X$ is consistently elongated compared to the ground state, whereas N$_\text{i}^*$-$X$ can either be elongated or shortened.  The variation of N$^*$-$X$ of amine N atoms (0.00--0.17 {\AA}) is larger than that of imine N atoms (0.00--0.06 {\AA}).  All N$_\text{a}^*$-H bond lengths are fixed at 0.97--0.98 {\AA}, which is reduced by 0.02--0.04 {\AA} compared to the GS geometry.  Meanwhile, we observed a consistent increase in the bond angle $\angle$C-N$_\text{i}^*$-$X$ upon ionization of the imine N atoms. The increment ranges over 3.2--3.6$^\circ$ when $X$ represents carbon and 2.4--2.9$^\circ$ when $X$ represents nitrogen. However, for amine N atoms, the bond angle $\angle$$X$-N$_\text{a}^*$-$X$  decreases for almost all molecules, except for the non-planar molecules 2-pyrroline, 3-pyrroline, and N1 in 2-pyrazoline. 

To reflect the global structural change, the root mean squared deviation (RMSD) values between the two structures were calculated and listed in Table \ref{tab:st}. Generally, amine nitrogens (0.03--0.15 {\AA})  give larger RMSD values than imine nitrogens (0.03--0.04 {\AA}). 

%---------------------------------------------------------------------------------
%+++++++++++++++++++++++++++++++
\subsection{Huang-Rhys factors ($S_{i}$) and the vibrational reorganization energies ($E_\text{r}$)}
%+++++++++++++++++++++++++++++++

Table \ref{si} presents the computed vibrational frequencies and Huang-Rhys factors (HRFs) for all molecules in their N1s ionized states. The HRF is defined as follows:
\begin{equation}
S_i=\frac{1}{2\hbar}\omega_i k_i^2.
\label{eq:Huang}
\end{equation}
Here, $k_i$  denotes the $i$-th elements of the displacement column vector $\mathbf{k}$, and  $\omega_i$  represents the vibrational frequency of mode $i$ in excited-state. The square root of the HRF, 
\begin{equation}
\lambda_i =\sqrt{S_{i}},
\end{equation}
is commonly referred as the electron-phonon coupling strength. In our analysis, we employed a threshold of $S_\text{i}$ $\geq$ 0.3 for all N$^*$ atoms.  Notably, we observed that only weak electron-vibration coupling ($S_\text{i}$ $\textless$ 1.0) was evident across all modes for all imine nitrogens.  In contrast, amine nitrogens exhibited modes with intermediate ($S_\text{i}$ $\approx$ 1.0)  or strong ($1.0 \le S_\text{i}\le 2.2$) electron-phonon coupling strengths.

We also computed the total vibrational reorganization energy, given by
\begin{equation}
E_\text{r}=\sum\limits_{i} S_\text{i} \hbar\omega_{i},
\end{equation}
for each molecule, which is closely associated with structural changes. Our calculations revealed that amine nitrogens (0.51--0.63 eV) exhibit significantly higher reorganization energy compared to imine nitrogens (0.10--0.34 eV). This indicates that ionization in amine nitrogen atoms induces more pronounced structural deformation than in imine nitrogen atoms.\cite{atahan-evrenk_quantitative_2018} Consequently, it is reasonable to observe smaller RMSDs for imine nitrogens when compared to amine nitrogens (as discussed in Section \ref{sec:rmsd} above).

%+++++++++++++++++++++++++++++++
\subsection{Vibronic fine structures}
%+++++++++++++++++++++++++++++++

\begin{figure*}
\includegraphics[width=0.5\textwidth]{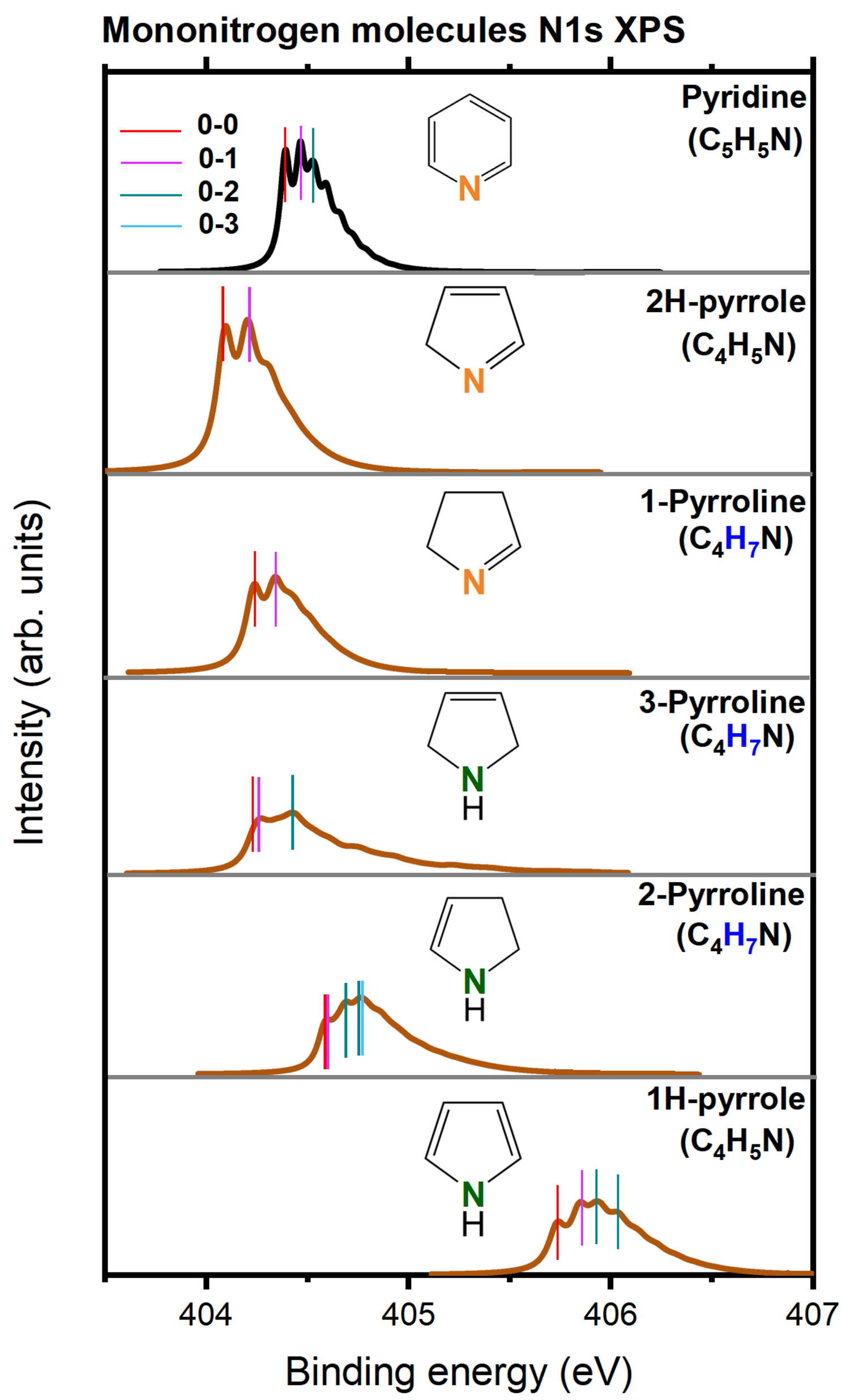}% Here is how to import EPS art
\caption{Simulated vibrationally-resolved N1s XPS spectra of mononitrogen molecules (2H-pyrrole, 1-Pyrroline, 3-Pyrroline, 2-Pyrroline, and 1H-pyrrole) using the FCH-DR method.  The amine and imine nitrogens contained in the molecule are highlighted in green and orange, respectively.  The pyridine spectrum\cite{wei_vibronic_2022} appeared at the top (black line) is shown for comparison.
}
\label{1n}
\end{figure*}

\begin{figure*}
\includegraphics[width=0.5\textwidth]{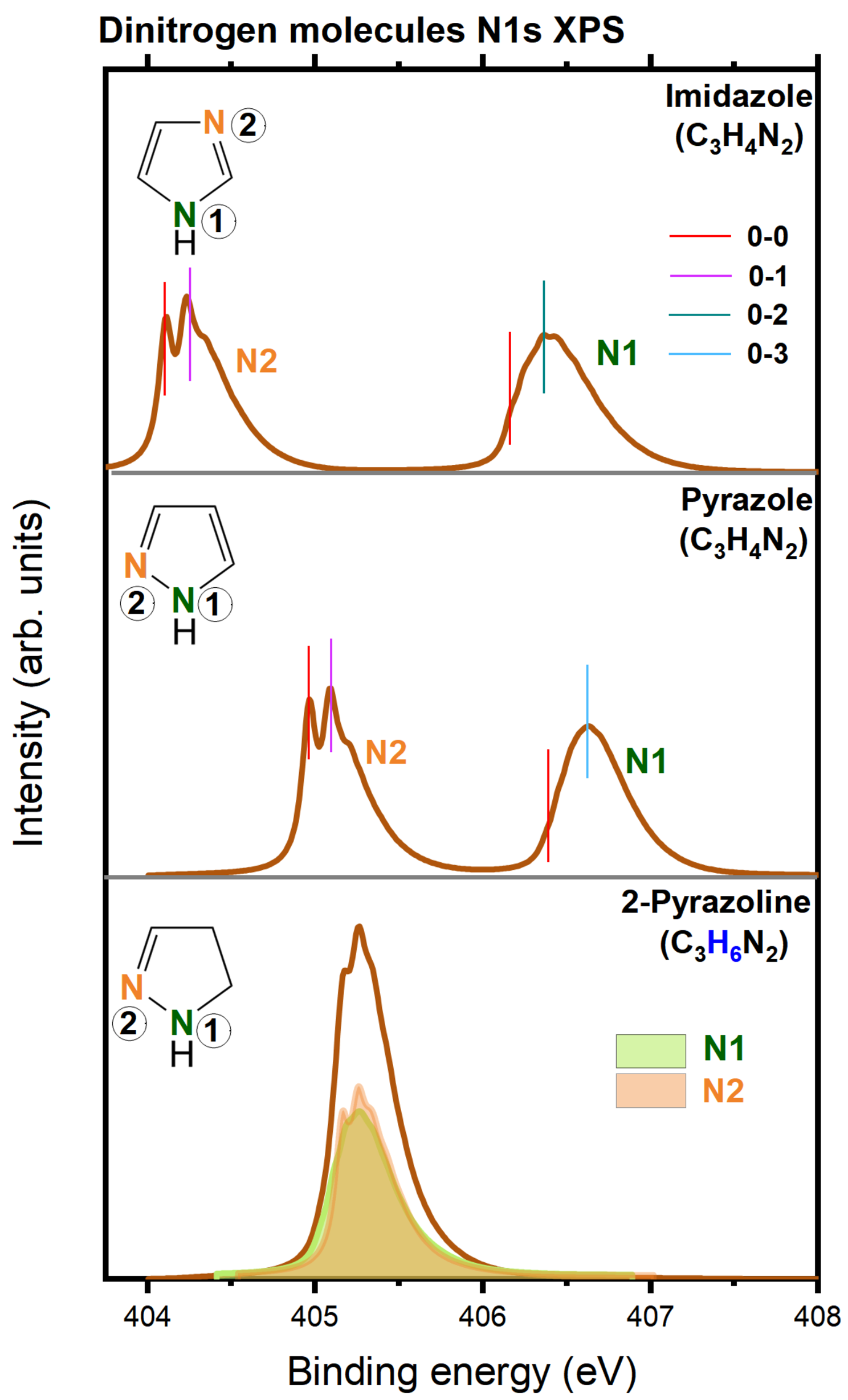}% Here is how to import EPS art
\caption{Simulated vibrationally-resolved  N1s XPS spectra  of imidazole, pyrazole, and 2-pyrazoline.
}
\label{2n}
\end{figure*}

\begin{figure*}
\includegraphics[width=0.5\textwidth]{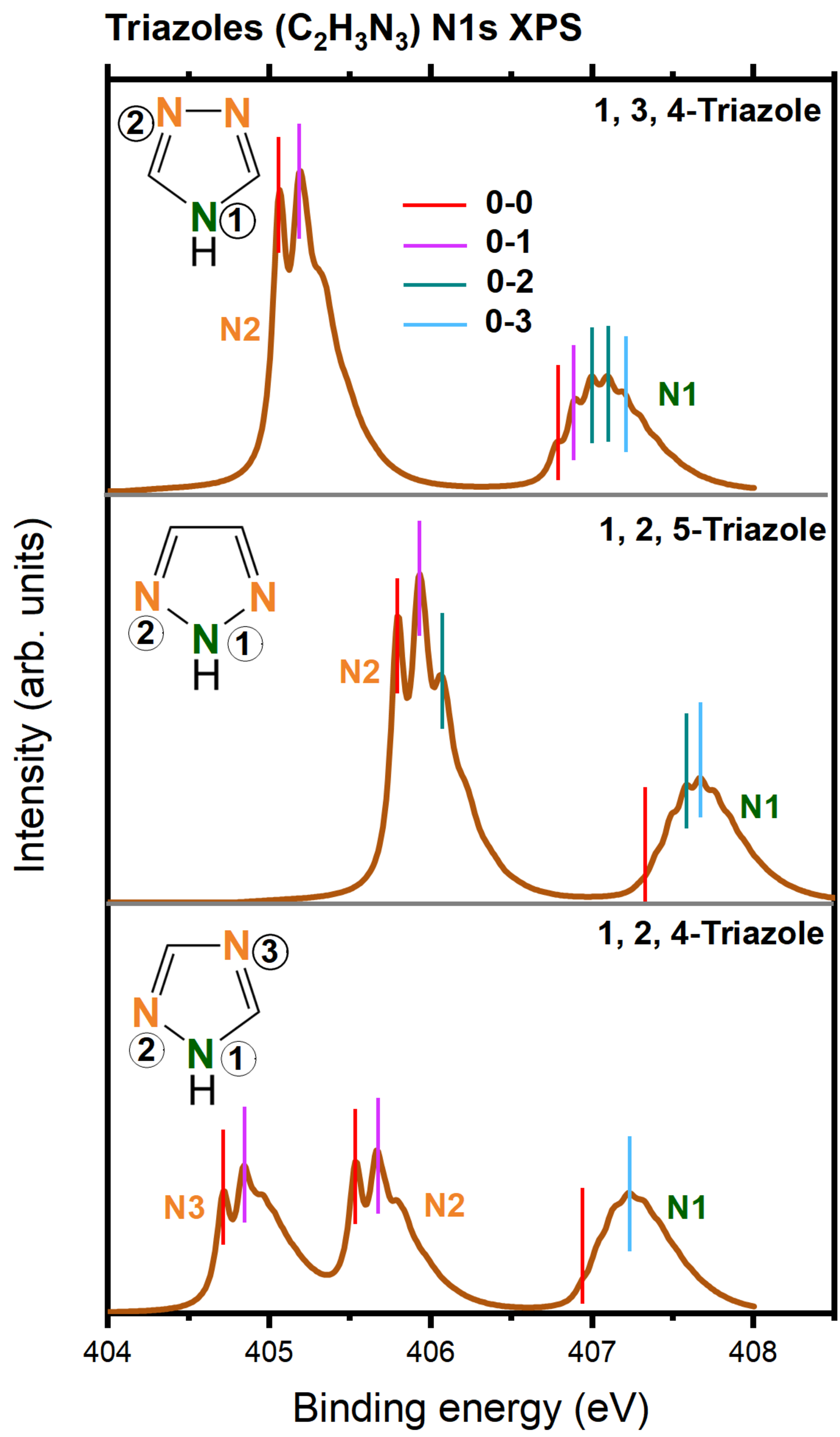}% Here is how to import EPS art
\caption{Comparison of the simulated  vibrationally-resolved N1s XPS spectra of 1,3,4-, 1,2,5-, and 1,2,4-triazoles.
}
\label{3n}
\end{figure*}

Figures \ref{1n} to \ref{3n} present the simulated vibrationally-resolved N1s XPS spectra of all molecules, highlighting the distinct differences in binding energies of different core holes. With the exception of 2-pyrroline, all polynitrogen molecules exhibit well-separated spectral contributions from N1 and N2 (1,2,4-triazole shows well-separated contributions from N1, N2, and N3). Figures S2 to S13 display the atom-specific spectra of all molecules, revealing significant variations between the spectra of imine and amine nitrogens. The sum of FC factors for imine nitrogens rapidly converges to 0.99 at $n$=5, whereas amine nitrogens require at least $n$=6. The spectra of all imine nitrogens display two distinct characteristic peaks similar to pyridine (top of Fig. \ref{1n}), primarily resulting from the 0-0 and 0-1 transitions (Figs. S2 to S13). In contrast, the spectra of amine nitrogens generally do not show a predominant 0-1 transition due to the significant mixing of various 0-$n$ transitions. This often leads to multiple oscillation-like weak fingerprints within a broad peak, which exhibit less pronounced spectral features when observed at low resolution.

%+++++++++++++++++++++++++++++++
\subsection{Effects of hydrogenation}
%+++++++++++++++++++++++++++++++

Hydrogenation can lead to significant changes in vertical IPs, as shown in three pairs of examples in Fig. S14. Amine nitrogens experience a decrease in IPs of 1.2--1.4 eV, while imine nitrogens undergo an increase of approximately 0.2 eV in response to molecular hydrogenation. 

Although hydrogenation has a significant effect on the IPs of amine and imine nitrogens, it has little impact on the spectral profile. For example, the vibrational characteristics of 1-pyrroline (Fig. S2) and 2H-pyrrole (Fig. S6),  2-pyrroline (Fig. S3) and 1H-pyrrole (Fig. S5), and N1/N2 of pyrazole (Fig. S8) and 2-pyrazoline (Fig. S9) are similar.  

The hydrogenation does not affect the spectra of imine nitrogens but slows down the convergence of the spectra for amine nitrogens  (a convergence threshold of 0.99 is constantly used throughout the work).  For instance, the imine nitrogen spectra converge at $n$ = 5 for both 1-pyrroline and 2H-pyrrole.  N2 (imine) in pyrazole and 2-pyrazoline behave similarly.  However, for amine nitrogen, 2-pyrroline ($n$ = 9) is more difficult to converge than 1H-pyrrole ($n$ = 6).  Likewise, N1 (amine) of 2-pyrazoline and pyrazole converge at $n$ = 8 and $n$ = 7, respectively.
 
%---------------------------------------------------
\section{\label{sec:discussion} Discussion: General vibronic coupling rules in small N-heterocycles}
%---------------------------------------------------
%The previous section gives the results for five-membered N-heterocycles.  

In this section, we discuss general vibronic coupling rules based on results of papers I,\cite{wei_vibronic_2022} II,\cite{wei_vibronic_2023} and III, covering mono- and bicyclic N-heterocycles. Special focus was paid on the difference between amine and imine nitrogens.

%+++++++++++++++++++++++++++++++
\subsection{IPs} 
%+++++++++++++++++++++++++++++++ 

For all 35 N-heterocyclic molecules we have studied in this series, the vertical IPs lie within a range of 4.6 eV from 403.3 to 407.9 eV. The three families, namely five-membered rings (404.5--407.9 eV), six-membered rings (404.5--407.0 eV), and bicyclic rings (403.3--407.2 eV), exhibit similar energy ranges. The connected N$^*$-N bond tends to exhibit higher IPs for the N$^*$ core hole.\cite{wei_vibronic_2022, wei_vibronic_2023}  For example, the BEs of N1 (N2) exhibit relative chemical shifts of 0.2 eV (0.9 eV) for imidazole and pyrazole molecules with the meta- and ortho-positions of the two nitrogens.  The 1,3,4-triazole, 1,2,5-triazole, and 1,2,4-triazole molecules manifest the chemical environment around the N1 core hole as C-N$^*$-C, N-N$^*$-C, and N-N$^*$-N, with corresponding vertical (adiabatic) ionization potentials of 407.2 (406.9), 407.4 (407.0), and 407.9 (407.4) eV, yielding relative chemical shifts of 0 (0), 0.2 (0.2), and 0.6 (0.5) eV, respectively. Obviously, these results are all due to the higher electronegativity of N atoms.

\begin{figure*}
\includegraphics[width=0.8\textwidth]{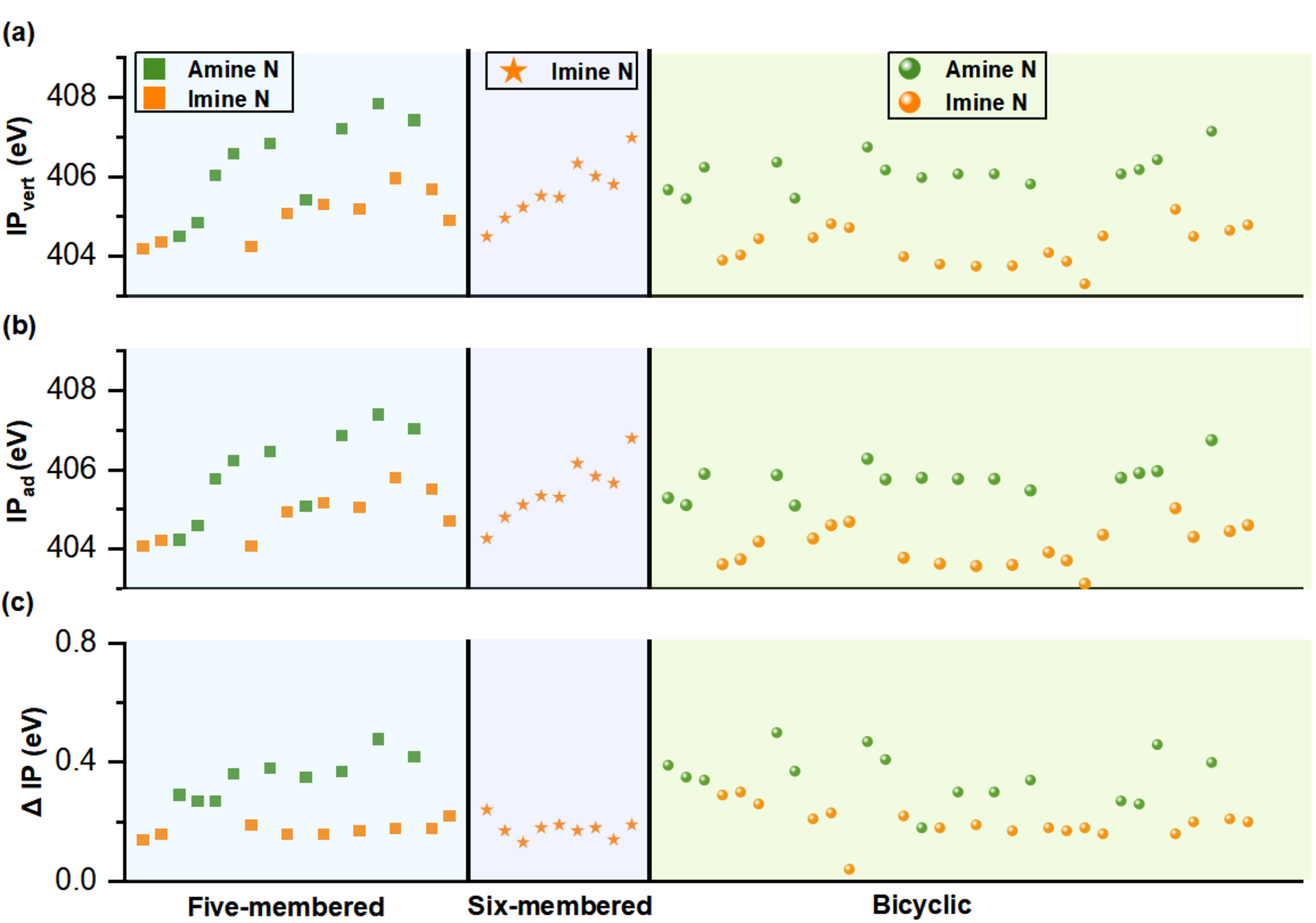}% Here is how to import EPS art
\caption{Simulated (a) vertical ($I^\text{vert}$), (b) adiabatic ($I^\text{ad}$) ionization potentials, and (c) their difference ($\Delta I$) of all 35 molecules studied in this series of papers [N-heterocycles with five-membered rings (this work), six-membered rings,\cite{wei_vibronic_2022} and bicyclic rings\cite{wei_vibronic_2023}].  Amine (green) and imine (orange) nitrogens are distinguished in color.}
\label{zong_ip}
\end{figure*}

As depicted in Fig. \ref{zong_ip}, amine N atoms generally exhibit larger BEs than those of imine N atoms. With respect to $\Delta I$, values associated with amine N atoms generally range between 0.3--0.5 eV, substantially larger than those of imine N atoms, which lie within the 0.1--0.2 eV range.  Consequently, it can be inferred that the PES displacement pertaining to amine N atoms typically exceeds that of imine N atoms.

\subsection{ZPE changes ($\Delta {\varepsilon}_{0}$)}

\begin{figure*}
\includegraphics[width=0.6\textwidth]{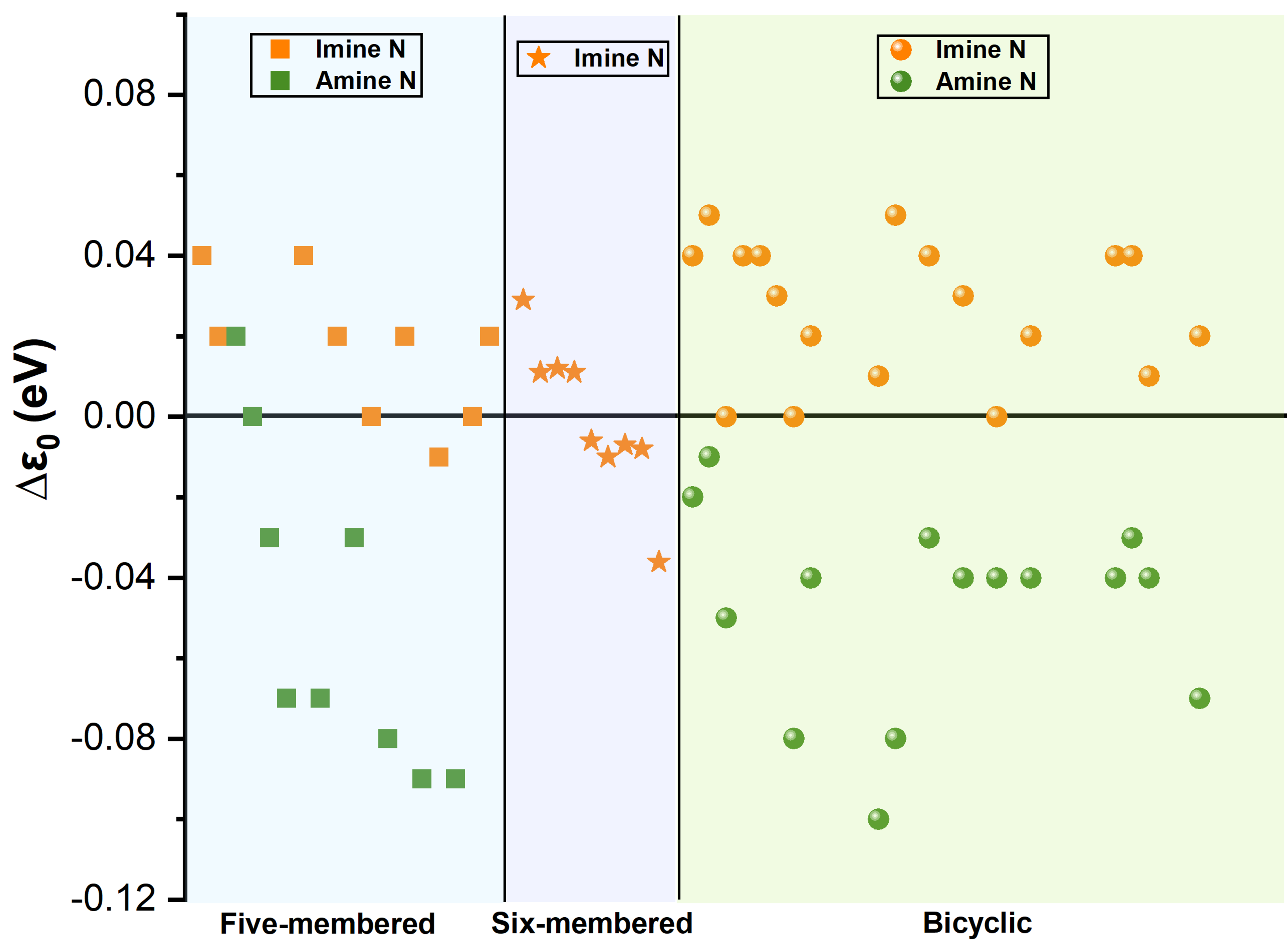}% Here is how to import EPS art
\caption{ Simulated $\Delta {\varepsilon}_{0}$ values of amine and imine nitrogens in all 35 molecules studied in this series of papers [N-heterocycles with five-membered rings (this work), six-membered rings,\cite{wei_vibronic_2022} and bicyclic rings\cite{wei_vibronic_2023}].  $\Delta {\varepsilon}_{0}$ of amine (green) and imine (orange) nitrogens are distinguished by colors.}
\label{zpe}
\end{figure*}

Figure \ref{zpe} shows the $\Delta {\varepsilon}_{0}$ values for each nonequivalent N in all 35 molecules. All 62 $\Delta {\varepsilon}_{0}$ values contributed from these nonequivalent nitrogen atoms cover a region of -0.10--0.05 eV, among which only 6 values are close to zero, indicating similar curvatures of PESs before and after the N1s ionization. For those nonzero values, it is interesting to find that most, if not all, imine nitrogens give positive values while amine nitrogens give negative values. This result indicates that the curvature generally becomes flatter (steeper) for amine (imine) N atoms. Nevertheless, exceptions indeed exist, particularly in six-membered N-heterocycles, where the positive and negative values are almost evenly distributed. 

\subsection{\label{sec:er_tol}Vibrational reorganization energies}

\begin{figure*}
\includegraphics[width=0.6\textwidth]{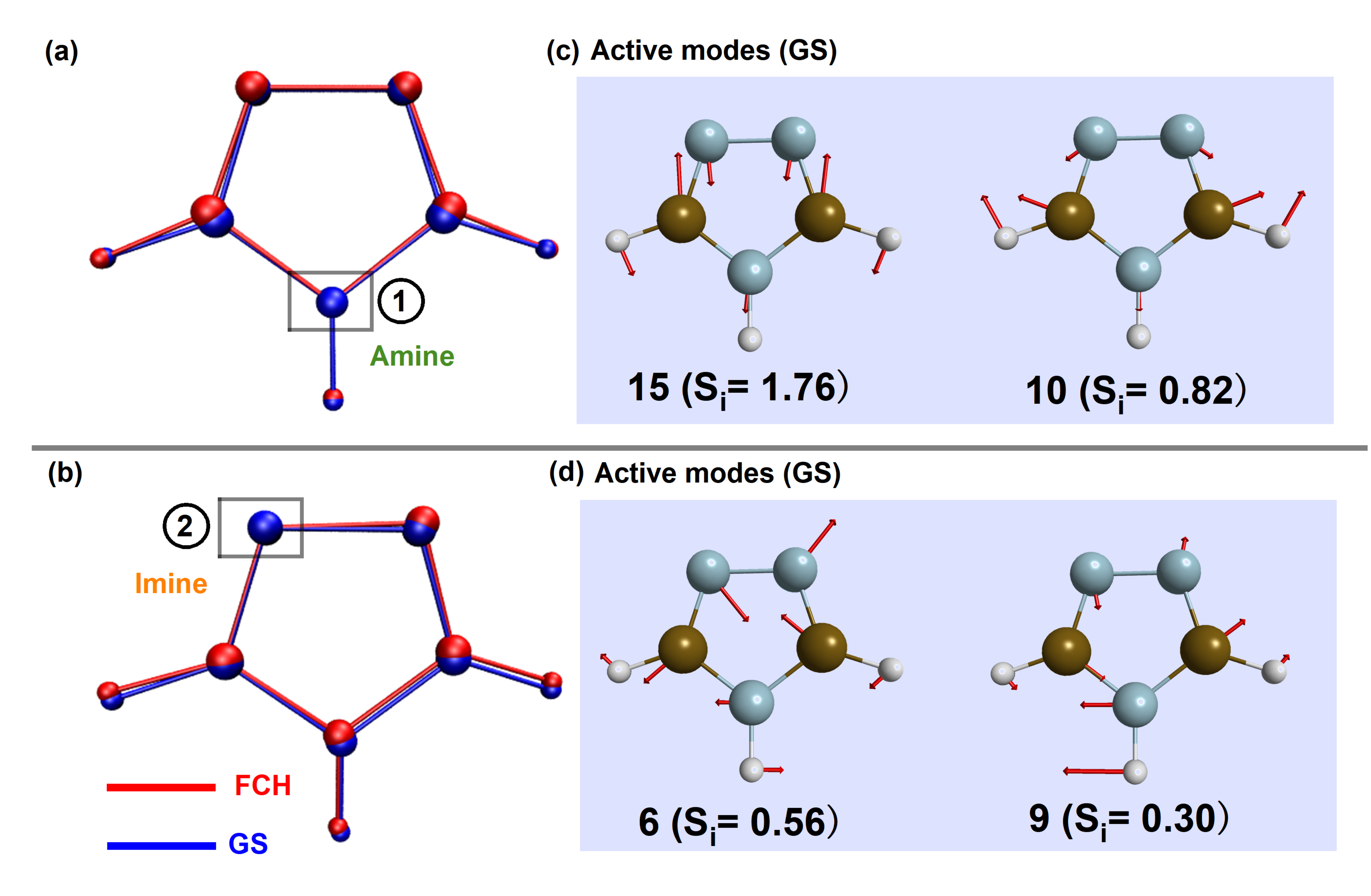}% Here is how to import EPS art
\caption{Simulated reorganization energies for all 35 molecules studied in this series of papers [N-heterocycles with five-membered rings (this work), six-membered rings,\cite{wei_vibronic_2022} and bicyclic rings\cite{wei_vibronic_2023}]. Amine (green) and imine (orange) nitrogen atoms are distinguished by color.}
\label{re}
\end{figure*}

The 1s ionization on amine and imine nitrogens leads to distinctly different structural changes.  The vibrational reorganization energies for each FCH state of all 35 molecules are listed in  Tables \ref{si}, S1, and S2, and are collectively visualized in Fig. \ref{re}.  The ionization of amine nitrogens (0.3--0.9 eV) produces greater reorganization energies than imine nitrogens (0.1--0.3 eV).  This indicates that amine nitrogens are more structurally altered than imine nitrogens during the N1s ionization process.

%+++++++++++++++++++++++++++++++
\subsection{Core ionization induced global and local structural changes}\label{jiegou}
%+++++++++++++++++++++++++++++++

The conclusion above for $E_\text{r}$ is consistent with the RMSD results between the ground state and the N1s ionized state structures, where amine nitrogens (0.03--0.15 \AA) generally exhibit larger RMSD values than imine nitrogens (0.03--0.05 \AA).

Apart from the global geometrical changes, we also summarize the local changes at the ionization center (N$^*$) for the series of molecules. The difference exists for nitrogens in five- and six-membered rings. For five-membered N-heterocycles, the bond length alternation (BLA) is consistent with the Kekul\'{e} structure, where the two N-$X$ distances at each N site exhibit the distinct tendency of one longer and one shorter distance.  In contrast, the BLA is more diminished for the six-membered ring structures, either in monocyclic \cite{wei_vibronic_2022} or bicyclic\cite{wei_vibronic_2023} molecules. That is because the difference between single and double bonds is effectively averaged within the conjugated $\pi$ bond system and resonating Kekul\'{e} structures. 
 
For the N$^*$--$X$ distances, amine N atoms (N$_\text{a}^*$--$X$, 1.40--1.56 {\AA}) generally have a longer distance than imine N atoms (N$_\text{i}^*$-$X$, 1.25--1.50 {\AA}).  The N$_\text{a}^*$-$X$ are always elongated, while N$_\text{i}^*$-$X$ can be either elongated or shortened. The N$_\text{a}^*$-H bond lengths in amine nitrogens remain nearly constant at 0.97--0.98 {\AA} and decrease only by 0.02--0.04 {\AA} compared to the GS geometry.
  
As for the bond angles $\angle$C-N$^*$-$X$, we found an increase for all imine N atoms ($\angle$C-N$_\text{i}^*$-$X$).  However, the bond angles $\angle$C-N$_\text{a}^*$-$X$ for amine N atoms are decreased for most molecules, except for three non-planar five-membered rings (2-pyrroline, 3-pyrroline, and N1 in 2-pyrazoline).

%+++++++++++++++++++++++++++++++
\subsection{Active modes}
%+++++++++++++++++++++++++++++++

The active ground-state vibrational modes of all 35 molecules are summarized in the Supplemental Material.\cite{si_five}  A threshold of HRF, $S_i\geq 0.3$, was adopted for all molecules. Within this standard, amine nitrogens usually exhibit more active vibrational modes (2--6 modes) compared to imine nitrogens (1--3 modes). For imine N1s ionizations, we can consistently identify a ring deformation mode that involves mainly N$^*$ and the atom (C or N) in its para-position.  Concerning amine nitrogens, another type of ring deformation mode is commonly identified, which involves N$^*$ along with ortho-position atoms (C or N).  No hydrogen vibrations (C-H stretching or bending) were found to be activated in the six-membered ring molecule. However, in both the five-membered and bicyclic molecules, active modes involving N/C-H bending are commonly found.

%+++++++++++++++++++++++++++++++
\subsection{Characteristics and interpretations of vibronic fine structures}
%+++++++++++++++++++++++++++++++

\begin{figure*}
\includegraphics[width=0.8\textwidth]{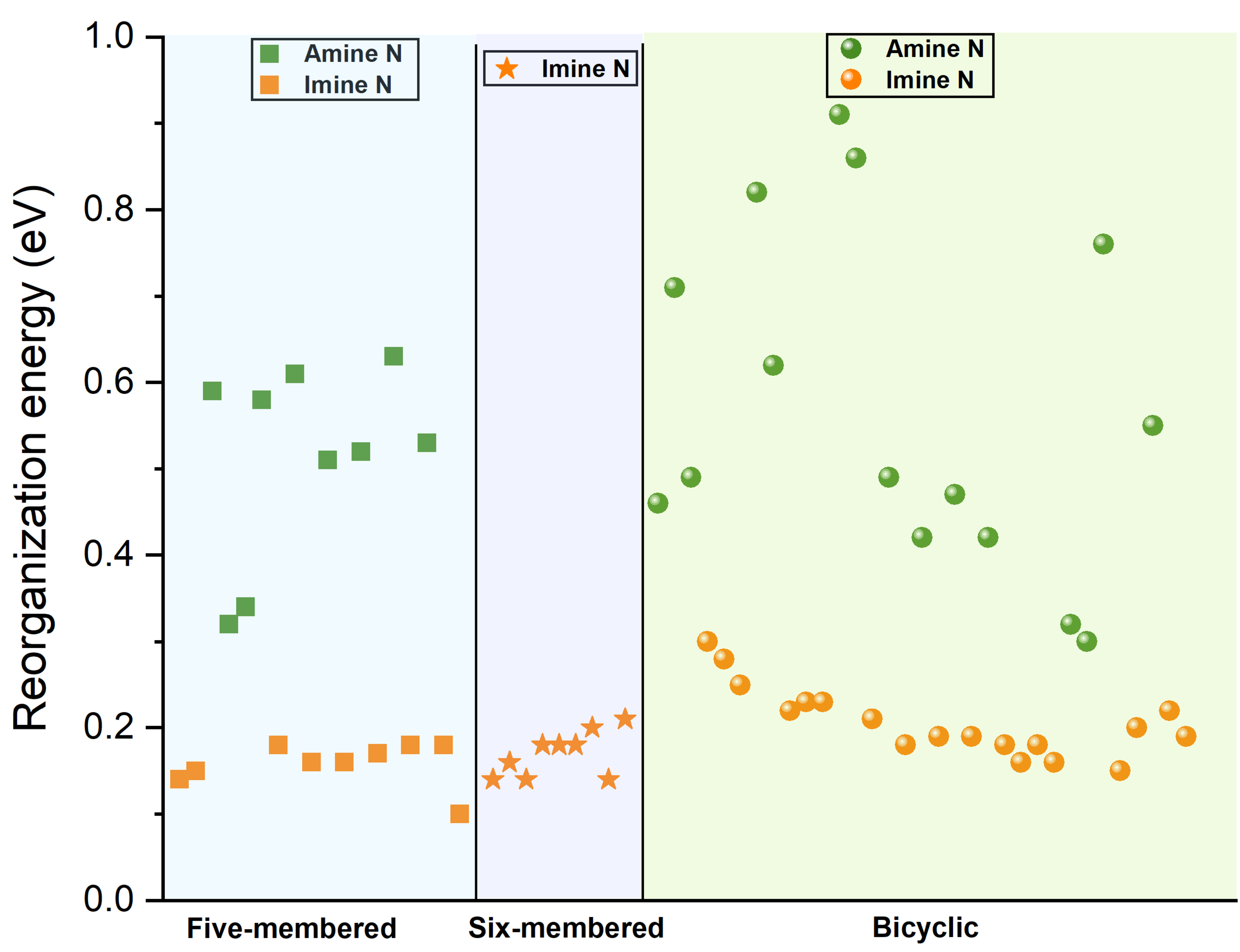}% Here is how to import EPS art
\caption{Analyses for  1,3,4-triazole. (a-b) Superposition of optimized geometries in the GS and FCH states: (a) N1 (amine N) and (b) N2 (imine N) 1s ionizations. Core ionized center is indicated by a black square. 
 (c-d) GS active vibrational modes. The number indicates the mode index with the Huang-Rhys factor given in parentheses. }
\label{134mode}
\end{figure*}

Generally, spectral profiles are related to the structural relaxation and vibrational coupling of the excited states.\cite{qiu_narrowband_2021} An increase in structural alterations between the ground and excited states results in a broadening of the spectral width. As discussed above, an amine nitrogen usually exhibits larger RMSD and reorganization energy than an imine nitrogen, leading to a broader spectral profile contributed by the amine nitrogen. For vibrational modes that exhibit strong coupling with structural changes, the high-order vibrational transition could have significant contributions and lead to larger Stokes shifts to the spectra.  Taking 1,3,4-triazole as an example, Fig. \ref{134mode} analyzes the correlation between the structural changes induced by core ionization and the ground-state active mode. In Fig. \ref{134mode}(a), we observe that N1 (amine nitrogen) 1s ionization leads to a significant shortening of the adjacent C-N bond length (from 1.30 to 1.25 \AA). This structural reorganization couples with modes 10 and 15, as depicted in Fig. \ref{134mode}(c), where mode 15 represents a strong electron-phonon coupling mode. Figure \ref{134mode}(b) shows that the prominent feature of the 1s ionization state of N2 (imine nitrogen) is the increase in the angle $\angle$C-N$^*$-N (from 107.4 to 110.3$^\circ$). Modes 6 and 9 couple with this structural change (corresponding HRFs are 0.56 and 0.30, respectively), but not as strongly as those observed for the case of N1. Consequently, the spectral signature of N2 (imine nitrogen) is dominated by the 0-1 transition, whereas the spectral profile of N1 (amine nitrogen) is dominated by the 0-$n$ transition with a relatively large $n$ ($n$=2).

%+++++++++++++++++++++++++++++++
\subsection{Effect of the benzene ring}
%+++++++++++++++++++++++++++++++

\begin{figure*}
\includegraphics[width=1.0\textwidth]{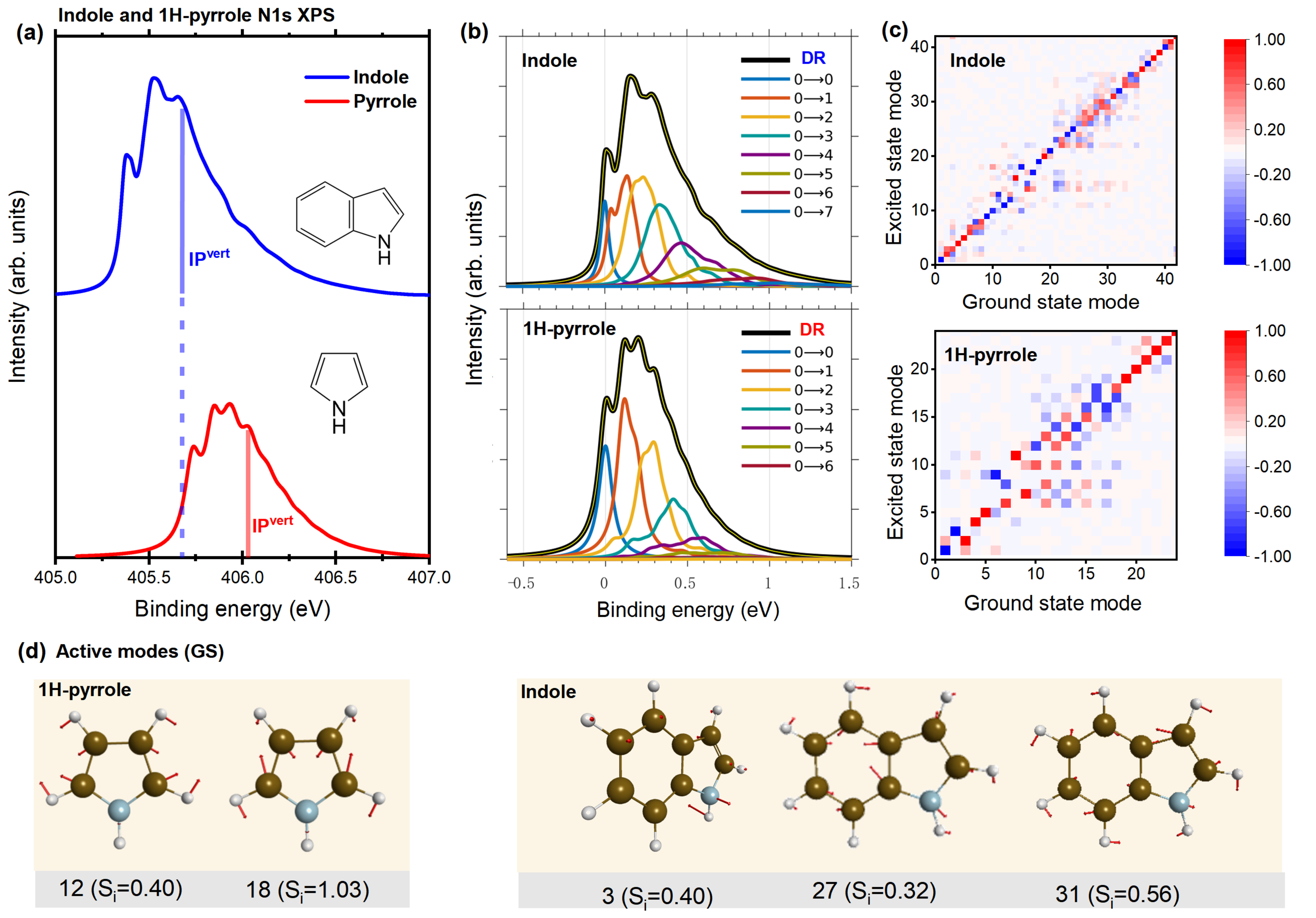}% Here is how to import EPS art
\caption{Comparison of indole and 1H-pyrrole based on (a) spectra (vertical ionization energies are indicated by vertical lines), (b) contributions of various 0-$n$ transitions to convergence, (c) the Duschinsky matrix, and (d)  active modes. 
}
\label{pyrrole+indole}
\end{figure*}

\begin{figure*}
\includegraphics[width=1.0\textwidth]{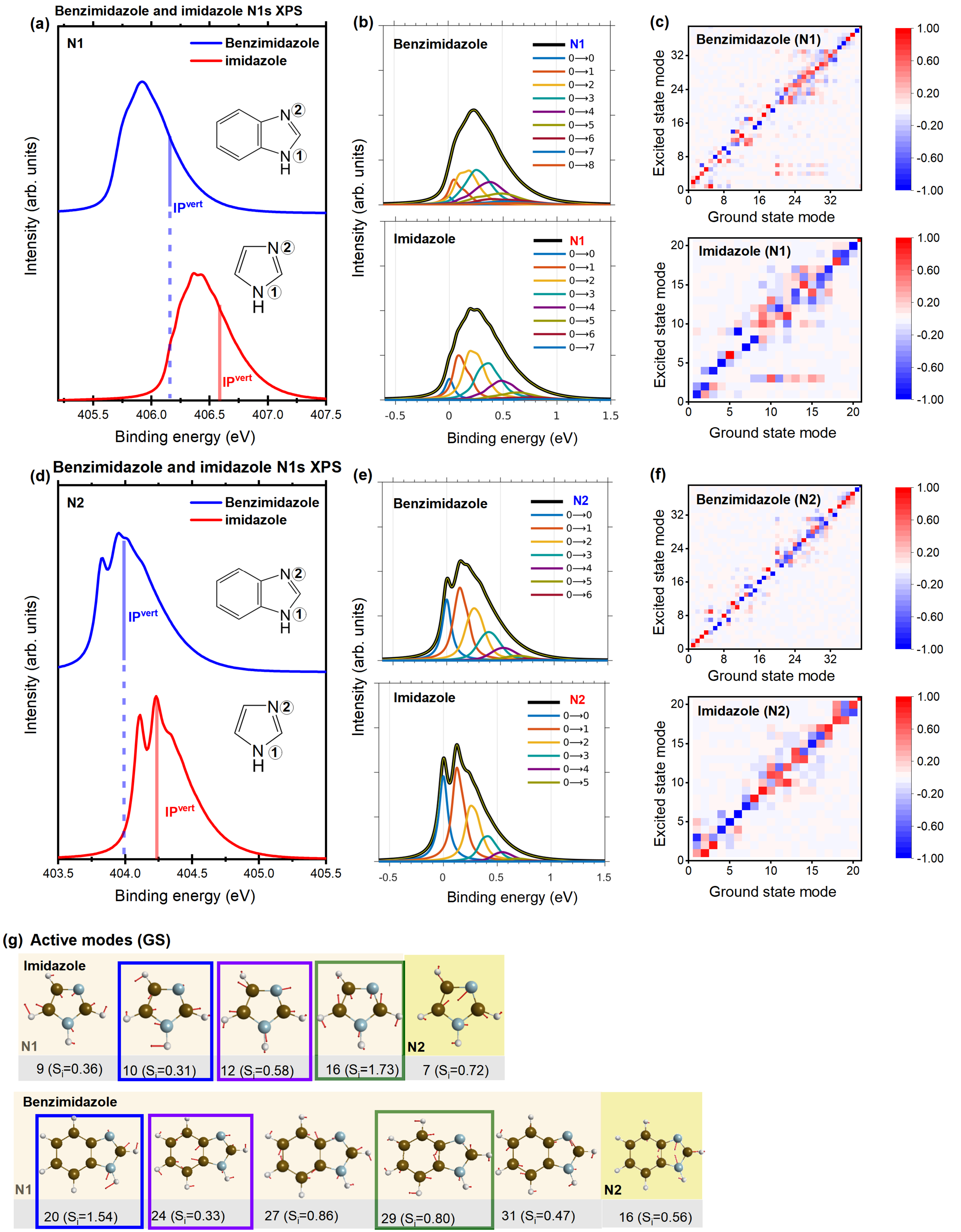}% Here is how to import EPS art
\caption{Comparison of benzimidazole and imidazole based on (a, d) spectra (vertical ionization energies are indicated by vertical lines), (b, e) contributions of the various 0-$n$ transitions to convergence, (c, f) the Duschinsky matrix, and (g)  active modes (colors surrounded by solid boxes represent different mode types).}
\label{ben+imi}
\end{figure*}

To examine the effect of the benzene ring, i.e. five-membered monocyclic and bicyclic molecules, three pairs of molecules are compared (pyrrole and indole in Fig. \ref{pyrrole+indole}, 1H-isoindole and 2H-pyrrole in Fig. {\isoindoleHpyrrole}, imidazole and benzimidazole in Fig. \ref{ben+imi}).  Within each pairing, the bicyclic molecules display a pronounced red shift in BEs relative to monocyclic molecules, attributable to the benzene ring acting as an electron-donating group (EDG).   Regarding spectra, five-membered rings invariably converge at a faster rate than their bicyclic counterparts, consequently resulting in narrower profiles. This phenomenon can be attributed to the lesser total reorganization energy of five-membered ring molecules, as depicted in Table \ref{erduibi}.  We examined the Duschinsky matrix for each pair of molecules and found that the bicyclic molecules displayed a more pronounced mode mixing effect [Fig. \ref{pyrrole+indole}(c), Fig. \ref{ben+imi}(c),(f), and Fig. {\isoindoleHpyrrole}(c)].

We have also conducted a comparison of the GS active modes for monocyclic and bicyclic molecules. As depicted in Figs. \ref{pyrrole+indole}(d), \ref{ben+imi}(g), and S15(d), each selected molecule features one active vibrational mode for the imine nitrogens, while amine nitrogens in bicyclic molecules have more active modes.  Although similar ring deformation modes are observed within the five-membered ring in both imidazole and benzimidazole, the vibrations of benzimidazole tend to be delocalized across both the five- and six-membered rings [Fig. \ref{ben+imi}(g)].  This result demonstrates that the introduction of a benzene ring can efficaciously alter the active modes.

\section{Conclusions and outlook}

In summary, we have simulated the vibrationally-resolved N1s XPS spectra of 12 five-membered N-heterocyclic compounds using the B3LYP-DR method and carried out a comprehensive analysis on the N 1s binding energies,  geometrical changes, major vibronic transitions, and active vibrational modes.  A clear distinction between amine and imine nitrogens was found, which is consistent with previous studies (paper I\cite{wei_vibronic_2022} and paper II\cite{wei_vibronic_2023}). 

Our series of investigations paints a comprehensive landscape for comprehending the N1s vibronic fine structure in small N-heterocycles.  For all 35 N-heterocycles in this series, the vibronic coupling rules are summarized. Distinctive performance of amine and imine nitrogens are summarized as follows. (1) Energies. Amine nitrogens generally have larger 1s BEs than imine nitrogens, and adjacent N-N bonds can increase the BEs. (2) Profiles. The spectra of imine nitrogens exhibit two distinct characteristic peaks originating from the 0-0 and 0-1 transitions, whereas the spectra of amine nitrogens are characterized by a broad peak with numerous weak fingerprints due to significant mixing of various 0-$n$ transitions. (3) PES.  N1s ionization generally leads to a pronounced alteration in the curvature direction of the final-state PES, which became flatter (steeper) for amine (imine) N atoms. The PES displacement associated with amine N atoms generally surpasses that of imine N atoms. (4) Reorganization energy. Amine nitrogens always have greater total vibrational reorganization energy than imine nitrogens. (5) Structure. Ionization for amine N atoms instigates more substantial global structural alterations compared to imine N atoms, evidenced by the difference between vertical and adiabatic ionization energies, the RMSD between \textbf{min GS} and \textbf{min FCH}, and reorganization energy. (6) The 1s ionization in amine N atoms always elongates N$^*$-C and shortens the N$^*$-H bond lengths, while both elongation and shortening in N$^*$-C bond lengths are observed for imine N atoms. Additionally, we also investigated the influence of benzene rings on vibronic coupling by comparing the properties of five-membered rings and bicyclic compounds.

%---------------------------------------------------------------------------------
\section{Acknowledgments}
%---------------------------------------------------------------------------------
Financial support from the National Natural Science Foundation of China (Grant No. 12274229)  is greatly acknowledged. M.W. thanks to Fund for Fostering Talented Doctoral Students of Nanjing University of Science and Technology.

%---------------------------------------------------------------------------------
%\bibliography{N1s}% Produces the bibliography via BibTeX.

\begin{thebibliography}{63}%
\makeatletter
\providecommand \@ifxundefined [1]{%
 \@ifx{#1\undefined}
}%
\providecommand \@ifnum [1]{%
 \ifnum #1\expandafter \@firstoftwo
 \else \expandafter \@secondoftwo
 \fi
}%
\providecommand \@ifx [1]{%
 \ifx #1\expandafter \@firstoftwo
 \else \expandafter \@secondoftwo
 \fi
}%
\providecommand \natexlab [1]{#1}%
\providecommand \enquote  [1]{``#1''}%
\providecommand \bibnamefont  [1]{#1}%
\providecommand \bibfnamefont [1]{#1}%
\providecommand \citenamefont [1]{#1}%
\providecommand \href@noop [0]{\@secondoftwo}%
\providecommand \href [0]{\begingroup \@sanitize@url \@href}%
\providecommand \@href[1]{\@@startlink{#1}\@@href}%
\providecommand \@@href[1]{\endgroup#1\@@endlink}%
\providecommand \@sanitize@url [0]{\catcode `\\12\catcode `\$12\catcode
  `\&12\catcode `\#12\catcode `\^12\catcode `\_12\catcode `\%12\relax}%
\providecommand \@@startlink[1]{}%
\providecommand \@@endlink[0]{}%
\providecommand \url  [0]{\begingroup\@sanitize@url \@url }%
\providecommand \@url [1]{\endgroup\@href {#1}{\urlprefix }}%
\providecommand \urlprefix  [0]{URL }%
\providecommand \Eprint [0]{\href }%
\providecommand \doibase [0]{https://doi.org/}%
\providecommand \selectlanguage [0]{\@gobble}%
\providecommand \bibinfo  [0]{\@secondoftwo}%
\providecommand \bibfield  [0]{\@secondoftwo}%
\providecommand \translation [1]{[#1]}%
\providecommand \BibitemOpen [0]{}%
\providecommand \bibitemStop [0]{}%
\providecommand \bibitemNoStop [0]{.\EOS\space}%
\providecommand \EOS [0]{\spacefactor3000\relax}%
\providecommand \BibitemShut  [1]{\csname bibitem#1\endcsname}%
\let\auto@bib@innerbib\@empty
%</preamble>
\bibitem [{\citenamefont {Rumble~Jr.}\ \emph {et~al.}(1992)\citenamefont
  {Rumble~Jr.}, \citenamefont {Bickham},\ and\ \citenamefont
  {Powell}}]{Rumble1992NIST}%
  \BibitemOpen
  \bibfield  {author} {\bibinfo {author} {\bibfnamefont {J.~R.}\ \bibnamefont
  {Rumble~Jr.}}, \bibinfo {author} {\bibfnamefont {D.~M.}\ \bibnamefont
  {Bickham}},\ and\ \bibinfo {author} {\bibfnamefont {C.~J.}\ \bibnamefont
  {Powell}},\ }\bibfield  {title} {\bibinfo {title} {The nist x-ray
  photoelectron spectroscopy database},\ }\href@noop {} {\bibfield  {journal}
  {\bibinfo  {journal} {Surf. Interface Anal.}\ }\textbf {\bibinfo {volume}
  {19}},\ \bibinfo {pages} {241} (\bibinfo {year} {1992})}\BibitemShut
  {NoStop}%
\bibitem [{\citenamefont {Jolly}\ \emph {et~al.}(1984)\citenamefont {Jolly},
  \citenamefont {Bomben},\ and\ \citenamefont
  {Eyermann}}]{jolly_core-electron_1984}%
  \BibitemOpen
  \bibfield  {author} {\bibinfo {author} {\bibfnamefont {W.}~\bibnamefont
  {Jolly}}, \bibinfo {author} {\bibfnamefont {K.}~\bibnamefont {Bomben}},\ and\
  \bibinfo {author} {\bibfnamefont {C.}~\bibnamefont {Eyermann}},\ }\bibfield
  {title} {\bibinfo {title} {Core-electron binding energies for gaseous atoms
  and molecules},\ }\href@noop {} {\bibfield  {journal} {\bibinfo  {journal}
  {Atomic Data and Nuclear Data Tables}\ }\textbf {\bibinfo {volume} {31}},\
  \bibinfo {pages} {433} (\bibinfo {year} {1984})}\BibitemShut {NoStop}%
\bibitem [{\citenamefont {Wagner}(1979)}]{wagner_1979_handbook}%
  \BibitemOpen
  \bibfield  {author} {\bibinfo {author} {\bibfnamefont {C.~D.}\ \bibnamefont
  {Wagner}},\ }\href@noop {} {\emph {\bibinfo {title} {Handbook of x-ray
  photoelectron spectroscopy: a reference book of standard data for use in
  x-ray photoelectron spectroscopy}}}\ (\bibinfo  {publisher} {Perkin-Elmer},\
  \bibinfo {year} {1979})\BibitemShut {NoStop}%
\bibitem [{int()}]{interxps}%
  \BibitemOpen
  \href@noop {} {}\bibinfo {note} {Https://www.xpsdata.com/, Accessed on
  2023-5-31.}\BibitemShut {Stop}%
\bibitem [{XPS()}]{XPSlibrary}%
  \BibitemOpen
  \href@noop {} {}\bibinfo {note} {Https://www.xpslibrary.com/, Accessed on
  2023-5-31.}\BibitemShut {Stop}%
\bibitem [{esp()}]{espectra}%
  \BibitemOpen
  \href@noop {} {}\bibinfo {note} {Https://espectra.aip.org/, Accessed on
  2023-5-31.}\BibitemShut {Stop}%
\bibitem [{sas()}]{sasj}%
  \BibitemOpen
  \href@noop {} {}\bibinfo {note} {Http://www.sasj.jp/, Accessed on
  2023-5-31.}\BibitemShut {Stop}%
\bibitem [{LaS()}]{LaSurface}%
  \BibitemOpen
  \href@noop {} {}\bibinfo {note} {Http://www.lasurface.com/xps/index, php
  Accessed on 2023-5-31.}\BibitemShut {Stop}%
\bibitem [{\citenamefont
  {Siegbahn}(1982{\natexlab{a}})}]{siegbahn_electron_1982}%
  \BibitemOpen
  \bibfield  {author} {\bibinfo {author} {\bibfnamefont {K.}~\bibnamefont
  {Siegbahn}},\ }\bibfield  {title} {\bibinfo {title} {Electron spectroscopy
  for atoms, molecules, and condensed matter},\ }\href@noop {} {\bibfield
  {journal} {\bibinfo  {journal} {Rev. Mod. Phys.}\ }\textbf {\bibinfo {volume}
  {54}},\ \bibinfo {pages} {709} (\bibinfo {year}
  {1982}{\natexlab{a}})}\BibitemShut {NoStop}%
\bibitem [{\citenamefont {Watts}(2003)}]{watts2003introduction}%
  \BibitemOpen
  \bibfield  {author} {\bibinfo {author} {\bibfnamefont {J.}~\bibnamefont
  {Watts}},\ }\bibfield  {title} {\bibinfo {title} {An introduction to surface
  analysis by {XPS} and {AES}},\ }\href@noop {} {\bibfield  {journal} {\bibinfo
   {journal} {Chapter}\ }\textbf {\bibinfo {volume} {5}},\ \bibinfo {pages}
  {149} (\bibinfo {year} {2003})}\BibitemShut {NoStop}%
\bibitem [{\citenamefont {Sokolowski}\ \emph {et~al.}(1957)\citenamefont
  {Sokolowski}, \citenamefont {Nordling},\ and\ \citenamefont
  {Siegbahn}}]{sokolowski1957magnetic}%
  \BibitemOpen
  \bibfield  {author} {\bibinfo {author} {\bibfnamefont {E.}~\bibnamefont
  {Sokolowski}}, \bibinfo {author} {\bibfnamefont {C.}~\bibnamefont
  {Nordling}},\ and\ \bibinfo {author} {\bibfnamefont {K.}~\bibnamefont
  {Siegbahn}},\ }\bibfield  {title} {\bibinfo {title} {Magnetic analysis of
  x-ray produced photo and auger electrons},\ }\href@noop {} {\bibfield
  {journal} {\bibinfo  {journal} {Arkiv Fysik}\ }\textbf {\bibinfo {volume}
  {12}} (\bibinfo {year} {1957})}\BibitemShut {NoStop}%
\bibitem [{\citenamefont
  {Siegbahn}(1982{\natexlab{b}})}]{siegbahn1982electron}%
  \BibitemOpen
  \bibfield  {author} {\bibinfo {author} {\bibfnamefont {K.}~\bibnamefont
  {Siegbahn}},\ }\bibfield  {title} {\bibinfo {title} {Electron spectroscopy
  for atoms, molecules, and condensed matter},\ }\href@noop {} {\bibfield
  {journal} {\bibinfo  {journal} {Rev. Mod. Phys.}\ }\textbf {\bibinfo {volume}
  {54}},\ \bibinfo {pages} {709} (\bibinfo {year}
  {1982}{\natexlab{b}})}\BibitemShut {NoStop}%
\bibitem [{\citenamefont {Siegbahn}\ \emph {et~al.}(1967)\citenamefont
  {Siegbahn}, \citenamefont {Nordling}, \citenamefont {Fahlman}, \citenamefont
  {Nordberg}, \citenamefont {Hamrin}, \citenamefont {Hedman}, \citenamefont
  {Johansson}, \citenamefont {Bergmark}, \citenamefont {Karlsson},
  \citenamefont {Lindgren} \emph {et~al.}}]{siegbahn1967atomic}%
  \BibitemOpen
  \bibfield  {author} {\bibinfo {author} {\bibfnamefont {K.}~\bibnamefont
  {Siegbahn}}, \bibinfo {author} {\bibfnamefont {C.}~\bibnamefont {Nordling}},
  \bibinfo {author} {\bibfnamefont {A.}~\bibnamefont {Fahlman}}, \bibinfo
  {author} {\bibfnamefont {R.}~\bibnamefont {Nordberg}}, \bibinfo {author}
  {\bibfnamefont {K.}~\bibnamefont {Hamrin}}, \bibinfo {author} {\bibfnamefont
  {J.}~\bibnamefont {Hedman}}, \bibinfo {author} {\bibfnamefont
  {G.}~\bibnamefont {Johansson}}, \bibinfo {author} {\bibfnamefont
  {T.}~\bibnamefont {Bergmark}}, \bibinfo {author} {\bibfnamefont
  {S.}~\bibnamefont {Karlsson}}, \bibinfo {author} {\bibfnamefont
  {I.}~\bibnamefont {Lindgren}}, \emph {et~al.},\ }\href@noop {} {\emph
  {\bibinfo {title} {{ESCA}: atomic, molecular and solid state structure
  studies by means of electron spectroscopy}}}\ (\bibinfo  {publisher}
  {Almqvist \& Wiksell, Uppsala, Sweden},\ \bibinfo {year} {1967})\BibitemShut
  {NoStop}%
\bibitem [{\citenamefont {Siegbahn}\ \emph {et~al.}(1974)\citenamefont
  {Siegbahn}, \citenamefont {Allison},\ and\ \citenamefont
  {Allison}}]{siegbahn1974esca}%
  \BibitemOpen
  \bibfield  {author} {\bibinfo {author} {\bibfnamefont {K.}~\bibnamefont
  {Siegbahn}}, \bibinfo {author} {\bibfnamefont {D.~A.}\ \bibnamefont
  {Allison}},\ and\ \bibinfo {author} {\bibfnamefont {J.~H.}\ \bibnamefont
  {Allison}},\ }\href@noop {} {\emph {\bibinfo {title} {ESCA-photoelectron
  spectroscopy}}}\ (\bibinfo  {publisher} {CRC Press, Cleveland, Ohio},\
  \bibinfo {year} {1974})\BibitemShut {NoStop}%
\bibitem [{\citenamefont {Allison}\ \emph {et~al.}(1972)\citenamefont
  {Allison}, \citenamefont {Johansson}, \citenamefont {Allan}, \citenamefont
  {Gelius}, \citenamefont {Siegbahn}, \citenamefont {Allison},\ and\
  \citenamefont {Siegbahn}}]{allison_molecular_1972}%
  \BibitemOpen
  \bibfield  {author} {\bibinfo {author} {\bibfnamefont {D.}~\bibnamefont
  {Allison}}, \bibinfo {author} {\bibfnamefont {G.}~\bibnamefont {Johansson}},
  \bibinfo {author} {\bibfnamefont {C.}~\bibnamefont {Allan}}, \bibinfo
  {author} {\bibfnamefont {U.}~\bibnamefont {Gelius}}, \bibinfo {author}
  {\bibfnamefont {H.}~\bibnamefont {Siegbahn}}, \bibinfo {author}
  {\bibfnamefont {J.}~\bibnamefont {Allison}},\ and\ \bibinfo {author}
  {\bibfnamefont {K.}~\bibnamefont {Siegbahn}},\ }\bibfield  {title} {\bibinfo
  {title} {Molecular spectroscopy by means of {ESCA}},\ }\href@noop {}
  {\bibfield  {journal} {\bibinfo  {journal} {J. Electron. Spectrosc. Relat.
  Phenom.}\ }\textbf {\bibinfo {volume} {1}},\ \bibinfo {pages} {269} (\bibinfo
  {year} {1972})}\BibitemShut {NoStop}%
\bibitem [{\citenamefont {Gelius}\ \emph {et~al.}(1970)\citenamefont {Gelius},
  \citenamefont {Hed{\'e}n}, \citenamefont {Hedman}, \citenamefont {Lindberg},
  \citenamefont {Manne}, \citenamefont {Nordberg}, \citenamefont {Nordling},\
  and\ \citenamefont {Siegbahn}}]{gelius_molecular_1970}%
  \BibitemOpen
  \bibfield  {author} {\bibinfo {author} {\bibfnamefont {U.}~\bibnamefont
  {Gelius}}, \bibinfo {author} {\bibfnamefont {P.~F.}\ \bibnamefont
  {Hed{\'e}n}}, \bibinfo {author} {\bibfnamefont {J.}~\bibnamefont {Hedman}},
  \bibinfo {author} {\bibfnamefont {B.~J.}\ \bibnamefont {Lindberg}}, \bibinfo
  {author} {\bibfnamefont {R.}~\bibnamefont {Manne}}, \bibinfo {author}
  {\bibfnamefont {R.}~\bibnamefont {Nordberg}}, \bibinfo {author}
  {\bibfnamefont {C.}~\bibnamefont {Nordling}},\ and\ \bibinfo {author}
  {\bibfnamefont {K.}~\bibnamefont {Siegbahn}},\ }\bibfield  {title} {\bibinfo
  {title} {Molecular {Spectroscopy} by {Means} of {ESCA} {III}.
  \textit{{Carbon} compounds}},\ }\href@noop {} {\bibfield  {journal} {\bibinfo
   {journal} {Phys. Scr.}\ }\textbf {\bibinfo {volume} {2}},\ \bibinfo {pages}
  {70} (\bibinfo {year} {1970})}\BibitemShut {NoStop}%
\bibitem [{\citenamefont {Fellner-Feldegg}\ \emph {et~al.}(1975)\citenamefont
  {Fellner-Feldegg}, \citenamefont {Siegbahn}, \citenamefont {Asplund},
  \citenamefont {Kelfve},\ and\ \citenamefont
  {Siegbahn}}]{fellner-feldegg_esca_1975}%
  \BibitemOpen
  \bibfield  {author} {\bibinfo {author} {\bibfnamefont {H.}~\bibnamefont
  {Fellner-Feldegg}}, \bibinfo {author} {\bibfnamefont {H.}~\bibnamefont
  {Siegbahn}}, \bibinfo {author} {\bibfnamefont {L.}~\bibnamefont {Asplund}},
  \bibinfo {author} {\bibfnamefont {P.}~\bibnamefont {Kelfve}},\ and\ \bibinfo
  {author} {\bibfnamefont {K.}~\bibnamefont {Siegbahn}},\ }\bibfield  {title}
  {\bibinfo {title} {{ESCA} applied to liquids {IV}. {A} wire system for {ESCA}
  measurements on liquids},\ }\href@noop {} {\bibfield  {journal} {\bibinfo
  {journal} {J. Electron. Spectrosc. Relat. Phenom.}\ }\textbf {\bibinfo
  {volume} {7}},\ \bibinfo {pages} {421} (\bibinfo {year} {1975})}\BibitemShut
  {NoStop}%
\bibitem [{\citenamefont {Siegbarn}\ \emph {et~al.}(1974)\citenamefont
  {Siegbarn}, \citenamefont {Asplund}, \citenamefont {Kelfve}, \citenamefont
  {Hamrin}, \citenamefont {Karlsson},\ and\ \citenamefont
  {Siegbahn}}]{siegbarn_esca_1974}%
  \BibitemOpen
  \bibfield  {author} {\bibinfo {author} {\bibfnamefont {H.}~\bibnamefont
  {Siegbarn}}, \bibinfo {author} {\bibfnamefont {L.}~\bibnamefont {Asplund}},
  \bibinfo {author} {\bibfnamefont {P.}~\bibnamefont {Kelfve}}, \bibinfo
  {author} {\bibfnamefont {K.}~\bibnamefont {Hamrin}}, \bibinfo {author}
  {\bibfnamefont {L.}~\bibnamefont {Karlsson}},\ and\ \bibinfo {author}
  {\bibfnamefont {K.}~\bibnamefont {Siegbahn}},\ }\bibfield  {title} {\bibinfo
  {title} {{ESCA} applied to liquids. {II}. {Valence} and core electron spectra
  of formamide},\ }\href@noop {} {\bibfield  {journal} {\bibinfo  {journal} {J.
  Electron. Spectrosc. Relat. Phenom.}\ }\textbf {\bibinfo {volume} {5}},\
  \bibinfo {pages} {1059} (\bibinfo {year} {1974})}\BibitemShut {NoStop}%
\bibitem [{\citenamefont {Wei}\ \emph {et~al.}(2022)\citenamefont {Wei},
  \citenamefont {Cheng}, \citenamefont {Zhang}, \citenamefont {Zhang},
  \citenamefont {Wang}, \citenamefont {Ge}, \citenamefont {Tian},\ and\
  \citenamefont {Hua}}]{wei_vibronic_2022}%
  \BibitemOpen
  \bibfield  {author} {\bibinfo {author} {\bibfnamefont {M.}~\bibnamefont
  {Wei}}, \bibinfo {author} {\bibfnamefont {X.}~\bibnamefont {Cheng}}, \bibinfo
  {author} {\bibfnamefont {L.}~\bibnamefont {Zhang}}, \bibinfo {author}
  {\bibfnamefont {J.-R.}\ \bibnamefont {Zhang}}, \bibinfo {author}
  {\bibfnamefont {S.-Y.}\ \bibnamefont {Wang}}, \bibinfo {author}
  {\bibfnamefont {G.}~\bibnamefont {Ge}}, \bibinfo {author} {\bibfnamefont
  {G.}~\bibnamefont {Tian}},\ and\ \bibinfo {author} {\bibfnamefont
  {W.}~\bibnamefont {Hua}},\ }\bibfield  {title} {\bibinfo {title} {Vibronic
  fine structure in the nitrogen 1s photoelectron spectra of molecules from
  {Franck}-{Condon} simulations: {Azines}},\ }\href@noop {} {\bibfield
  {journal} {\bibinfo  {journal} {Phys. Rev. A}\ }\textbf {\bibinfo {volume}
  {106}},\ \bibinfo {pages} {022811} (\bibinfo {year} {2022})}\BibitemShut
  {NoStop}%
\bibitem [{\citenamefont {Wei}\ \emph {et~al.}(2023)\citenamefont {Wei},
  \citenamefont {Zhang}, \citenamefont {Tian},\ and\ \citenamefont
  {Hua}}]{wei_vibronic_2023}%
  \BibitemOpen
  \bibfield  {author} {\bibinfo {author} {\bibfnamefont {M.}~\bibnamefont
  {Wei}}, \bibinfo {author} {\bibfnamefont {L.}~\bibnamefont {Zhang}}, \bibinfo
  {author} {\bibfnamefont {G.}~\bibnamefont {Tian}},\ and\ \bibinfo {author}
  {\bibfnamefont {W.}~\bibnamefont {Hua}},\ }\bibfield  {title} {\bibinfo
  {title} {Vibronic fine structure in the nitrogen 1 s photoelectron spectra
  from {Franck}-{Condon} simulation. {II}. {Indoles}},\ }\href@noop {}
  {\bibfield  {journal} {\bibinfo  {journal} {Phys. Rev. A}\ }\textbf {\bibinfo
  {volume} {108}},\ \bibinfo {pages} {022816} (\bibinfo {year}
  {2023})}\BibitemShut {NoStop}%
\bibitem [{\citenamefont {Alongi}\ \emph {et~al.}(2005)\citenamefont {Alongi},
  \citenamefont {Minetto},\ and\ \citenamefont {Taddei}}]{alongi_new_2005}%
  \BibitemOpen
  \bibfield  {author} {\bibinfo {author} {\bibfnamefont {M.}~\bibnamefont
  {Alongi}}, \bibinfo {author} {\bibfnamefont {G.}~\bibnamefont {Minetto}},\
  and\ \bibinfo {author} {\bibfnamefont {M.}~\bibnamefont {Taddei}},\
  }\bibfield  {title} {\bibinfo {title} {New pyrrole-based amino acids for the
  synthesis of peptidomimetic constrained scaffolds},\ }\href@noop {}
  {\bibfield  {journal} {\bibinfo  {journal} {Tetrahedron Lett.}\ }\textbf
  {\bibinfo {volume} {46}},\ \bibinfo {pages} {7069} (\bibinfo {year}
  {2005})}\BibitemShut {NoStop}%
\bibitem [{\citenamefont {Wurtz}\ \emph {et~al.}(2001)\citenamefont {Wurtz},
  \citenamefont {Turner}, \citenamefont {Baird},\ and\ \citenamefont
  {Dervan}}]{wurtz_fmoc_2001}%
  \BibitemOpen
  \bibfield  {author} {\bibinfo {author} {\bibfnamefont {N.~R.}\ \bibnamefont
  {Wurtz}}, \bibinfo {author} {\bibfnamefont {J.~M.}\ \bibnamefont {Turner}},
  \bibinfo {author} {\bibfnamefont {E.~E.}\ \bibnamefont {Baird}},\ and\
  \bibinfo {author} {\bibfnamefont {P.~B.}\ \bibnamefont {Dervan}},\ }\bibfield
   {title} {\bibinfo {title} {Fmoc {Solid} {Phase} {Synthesis} of {Polyamides}
  {Containing} {Pyrrole} and {Imidazole} {Amino} {Acids}},\ }\href@noop {}
  {\bibfield  {journal} {\bibinfo  {journal} {Org. Lett.}\ }\textbf {\bibinfo
  {volume} {3}},\ \bibinfo {pages} {1201} (\bibinfo {year} {2001})}\BibitemShut
  {NoStop}%
\bibitem [{\citenamefont {Crossley}\ and\ \citenamefont
  {McDonald}(1999)}]{crossley_fused_1999}%
  \BibitemOpen
  \bibfield  {author} {\bibinfo {author} {\bibfnamefont {M.~J.}\ \bibnamefont
  {Crossley}}\ and\ \bibinfo {author} {\bibfnamefont {J.~A.}\ \bibnamefont
  {McDonald}},\ }\bibfield  {title} {\bibinfo {title} {Fused
  porphyrin-imidazole systems: new building blocks for synthesis of porphyrin
  arrays},\ }\href@noop {} {\bibfield  {journal} {\bibinfo  {journal} {J. Chem.
  Soc., Perkin Trans. 1}\ ,\ \bibinfo {pages} {2429}} (\bibinfo {year}
  {1999})}\BibitemShut {NoStop}%
\bibitem [{\citenamefont {Gomez}\ \emph {et~al.}(2013)\citenamefont {Gomez},
  \citenamefont {Betzer}, \citenamefont {Voituriez},\ and\ \citenamefont
  {Marinetti}}]{gomez_phosphine_2013}%
  \BibitemOpen
  \bibfield  {author} {\bibinfo {author} {\bibfnamefont {C.}~\bibnamefont
  {Gomez}}, \bibinfo {author} {\bibfnamefont {J.-F.}\ \bibnamefont {Betzer}},
  \bibinfo {author} {\bibfnamefont {A.}~\bibnamefont {Voituriez}},\ and\
  \bibinfo {author} {\bibfnamefont {A.}~\bibnamefont {Marinetti}},\ }\bibfield
  {title} {\bibinfo {title} {Phosphine {Organocatalysis} in the {Synthesis} of
  {Natural} {Products} and {Bioactive} {Compounds}},\ }\href@noop {} {\bibfield
   {journal} {\bibinfo  {journal} {ChemCatChem}\ }\textbf {\bibinfo {volume}
  {5}},\ \bibinfo {pages} {1055} (\bibinfo {year} {2013})}\BibitemShut
  {NoStop}%
\bibitem [{\citenamefont {Nguyen}\ \emph {et~al.}(2022)\citenamefont {Nguyen},
  \citenamefont {Dai}, \citenamefont {Mechler}, \citenamefont {Hoa},\ and\
  \citenamefont {Vo}}]{nguyen_synthesis_2022}%
  \BibitemOpen
  \bibfield  {author} {\bibinfo {author} {\bibfnamefont {N.~T.}\ \bibnamefont
  {Nguyen}}, \bibinfo {author} {\bibfnamefont {V.~V.}\ \bibnamefont {Dai}},
  \bibinfo {author} {\bibfnamefont {A.}~\bibnamefont {Mechler}}, \bibinfo
  {author} {\bibfnamefont {N.~T.}\ \bibnamefont {Hoa}},\ and\ \bibinfo {author}
  {\bibfnamefont {Q.~V.}\ \bibnamefont {Vo}},\ }\bibfield  {title} {\bibinfo
  {title} {Synthesis and evaluation of the antioxidant activity of
  3-pyrroline-2-ones: experimental and theoretical insights},\ }\href@noop {}
  {\bibfield  {journal} {\bibinfo  {journal} {RSC Adv.}\ }\textbf {\bibinfo
  {volume} {12}},\ \bibinfo {pages} {24579} (\bibinfo {year}
  {2022})}\BibitemShut {NoStop}%
\bibitem [{\citenamefont {Sun}\ \emph {et~al.}(2020)\citenamefont {Sun},
  \citenamefont {Zou}, \citenamefont {Wang},\ and\ \citenamefont
  {Yang}}]{sun_sulfur-directed_2020}%
  \BibitemOpen
  \bibfield  {author} {\bibinfo {author} {\bibfnamefont {G.}~\bibnamefont
  {Sun}}, \bibinfo {author} {\bibfnamefont {X.}~\bibnamefont {Zou}}, \bibinfo
  {author} {\bibfnamefont {J.}~\bibnamefont {Wang}},\ and\ \bibinfo {author}
  {\bibfnamefont {W.}~\bibnamefont {Yang}},\ }\bibfield  {title} {\bibinfo
  {title} {Sulfur-directed palladium-catalyzed
  {C}(sp$^{\textrm{3}}$){\textendash}{H} $\alpha$-arylation of 3-pyrrolines:
  easy access to diverse polysubstituted pyrrolidines},\ }\href@noop {}
  {\bibfield  {journal} {\bibinfo  {journal} {Org. Chem. Front.}\ }\textbf
  {\bibinfo {volume} {7}},\ \bibinfo {pages} {666} (\bibinfo {year}
  {2020})}\BibitemShut {NoStop}%
\bibitem [{\citenamefont {Anderson}\ and\ \citenamefont
  {Milowsky}(1987)}]{anderson_3-pyrroline_1987}%
  \BibitemOpen
  \bibfield  {author} {\bibinfo {author} {\bibfnamefont {W.~K.}\ \bibnamefont
  {Anderson}}\ and\ \bibinfo {author} {\bibfnamefont {A.~S.}\ \bibnamefont
  {Milowsky}},\ }\bibfield  {title} {\bibinfo {title} {3-{Pyrroline} {N}-oxide
  bis(carbamate) tumor inhibitors as analogs of indicine {N}-oxide},\
  }\href@noop {} {\bibfield  {journal} {\bibinfo  {journal} {J. Med. Chem.}\
  }\textbf {\bibinfo {volume} {30}},\ \bibinfo {pages} {2144} (\bibinfo {year}
  {1987})}\BibitemShut {NoStop}%
\bibitem [{\citenamefont {Chen}\ \emph {et~al.}(2020)\citenamefont {Chen},
  \citenamefont {Zhang}, \citenamefont {Lu}, \citenamefont {Wang},
  \citenamefont {Si}, \citenamefont {Yan},\ and\ \citenamefont
  {Yang}}]{chen_novel_2020}%
  \BibitemOpen
  \bibfield  {author} {\bibinfo {author} {\bibfnamefont {M.}~\bibnamefont
  {Chen}}, \bibinfo {author} {\bibfnamefont {L.}~\bibnamefont {Zhang}},
  \bibinfo {author} {\bibfnamefont {A.}~\bibnamefont {Lu}}, \bibinfo {author}
  {\bibfnamefont {X.}~\bibnamefont {Wang}}, \bibinfo {author} {\bibfnamefont
  {W.}~\bibnamefont {Si}}, \bibinfo {author} {\bibfnamefont {J.}~\bibnamefont
  {Yan}},\ and\ \bibinfo {author} {\bibfnamefont {C.}~\bibnamefont {Yang}},\
  }\bibfield  {title} {\bibinfo {title} {Novel carboxylated pyrroline-2-one
  derivatives bearing a phenylhydrazine moiety: {Design}, synthesis, antifungal
  evaluation and {3D}-{QSAR} analysis},\ }\href@noop {} {\bibfield  {journal}
  {\bibinfo  {journal} {Bioorg. Med. Chem. Lett.}\ }\textbf {\bibinfo {volume}
  {30}},\ \bibinfo {pages} {127519} (\bibinfo {year} {2020})}\BibitemShut
  {NoStop}%
\bibitem [{\citenamefont {Kost}\ and\ \citenamefont
  {Grandberg}(1966)}]{kost_progress_1966}%
  \BibitemOpen
  \bibfield  {author} {\bibinfo {author} {\bibfnamefont {A.}~\bibnamefont
  {Kost}}\ and\ \bibinfo {author} {\bibfnamefont {I.}~\bibnamefont
  {Grandberg}},\ }\bibfield  {title} {\bibinfo {title} {Progress in {Pyrazole}
  {Chemistry}},\ }in\ \href@noop {} {\emph {\bibinfo {booktitle} {Adv.
  Heterocycl. Chem.}}},\ Vol.~\bibinfo {volume} {6}\ (\bibinfo  {publisher}
  {Elsevier},\ \bibinfo {year} {1966})\ pp.\ \bibinfo {pages}
  {347--429}\BibitemShut {NoStop}%
\bibitem [{\citenamefont {Dippold}\ and\ \citenamefont
  {Klap{\"o}tke}(2013)}]{dippold_study_2013}%
  \BibitemOpen
  \bibfield  {author} {\bibinfo {author} {\bibfnamefont {A.~A.}\ \bibnamefont
  {Dippold}}\ and\ \bibinfo {author} {\bibfnamefont {T.~M.}\ \bibnamefont
  {Klap{\"o}tke}},\ }\bibfield  {title} {\bibinfo {title} {A {Study} of
  {Dinitro}-bis-1,2,4-triazole-1,1'-diol and {Derivatives}: {Design} of
  {High}-{Performance} {Insensitive} {Energetic} {Materials} by the
  {Introduction} of {N}-{Oxides}},\ }\href@noop {} {\bibfield  {journal}
  {\bibinfo  {journal} {J. Am. Chem. Soc.}\ }\textbf {\bibinfo {volume}
  {135}},\ \bibinfo {pages} {9931} (\bibinfo {year} {2013})}\BibitemShut
  {NoStop}%
\bibitem [{\citenamefont {Ma}\ \emph {et~al.}(2018)\citenamefont {Ma},
  \citenamefont {Cheng}, \citenamefont {Ju}, \citenamefont {Yi}, \citenamefont
  {Zhu}, \citenamefont {Zhang},\ and\ \citenamefont
  {Yang}}]{ma_amino-nitramino_2018}%
  \BibitemOpen
  \bibfield  {author} {\bibinfo {author} {\bibfnamefont {J.}~\bibnamefont
  {Ma}}, \bibinfo {author} {\bibfnamefont {G.}~\bibnamefont {Cheng}}, \bibinfo
  {author} {\bibfnamefont {X.}~\bibnamefont {Ju}}, \bibinfo {author}
  {\bibfnamefont {Z.}~\bibnamefont {Yi}}, \bibinfo {author} {\bibfnamefont
  {S.}~\bibnamefont {Zhu}}, \bibinfo {author} {\bibfnamefont {Z.}~\bibnamefont
  {Zhang}},\ and\ \bibinfo {author} {\bibfnamefont {H.}~\bibnamefont {Yang}},\
  }\bibfield  {title} {\bibinfo {title} {Amino-nitramino functionalized
  triazolotriazines: a good balance between high energy and low sensitivity},\
  }\href@noop {} {\bibfield  {journal} {\bibinfo  {journal} {Dalton Trans.}\
  }\textbf {\bibinfo {volume} {47}},\ \bibinfo {pages} {14483} (\bibinfo {year}
  {2018})}\BibitemShut {NoStop}%
\bibitem [{\citenamefont {Wang}\ \emph {et~al.}(2018)\citenamefont {Wang},
  \citenamefont {Xu}, \citenamefont {Lin},\ and\ \citenamefont
  {Lu}}]{wang_recent_2018}%
  \BibitemOpen
  \bibfield  {author} {\bibinfo {author} {\bibfnamefont {P.}~\bibnamefont
  {Wang}}, \bibinfo {author} {\bibfnamefont {Y.}~\bibnamefont {Xu}}, \bibinfo
  {author} {\bibfnamefont {Q.}~\bibnamefont {Lin}},\ and\ \bibinfo {author}
  {\bibfnamefont {M.}~\bibnamefont {Lu}},\ }\bibfield  {title} {\bibinfo
  {title} {Recent advances in the syntheses and properties of polynitrogen
  pentazolate anion \textit{cyclo}-{N}$_{\textrm{5}}$$^{\textrm{-}}$ and its
  derivatives},\ }\href@noop {} {\bibfield  {journal} {\bibinfo  {journal}
  {Chem. Soc. Rev.}\ }\textbf {\bibinfo {volume} {47}},\ \bibinfo {pages}
  {7522} (\bibinfo {year} {2018})}\BibitemShut {NoStop}%
\bibitem [{\citenamefont {Zhang}\ \emph {et~al.}(2017)\citenamefont {Zhang},
  \citenamefont {Sun}, \citenamefont {Hu}, \citenamefont {Yu},\ and\
  \citenamefont {Lu}}]{zhang_synthesis_2017}%
  \BibitemOpen
  \bibfield  {author} {\bibinfo {author} {\bibfnamefont {C.}~\bibnamefont
  {Zhang}}, \bibinfo {author} {\bibfnamefont {C.}~\bibnamefont {Sun}}, \bibinfo
  {author} {\bibfnamefont {B.}~\bibnamefont {Hu}}, \bibinfo {author}
  {\bibfnamefont {C.}~\bibnamefont {Yu}},\ and\ \bibinfo {author}
  {\bibfnamefont {M.}~\bibnamefont {Lu}},\ }\bibfield  {title} {\bibinfo
  {title} {Synthesis and characterization of the pentazolate anion
  \textit{cyclo}-{N}$_5^-$ in
  ({N}$_{\textrm{5}}$)$_{\textrm{6}}$({H}$_{\textrm{3}}${O})$_{\textrm{3}}$({NH}$_{\textrm{4}}$)$_{\textrm{4}}${Cl}},\
  }\href@noop {} {\bibfield  {journal} {\bibinfo  {journal} {Science}\ }\textbf
  {\bibinfo {volume} {355}},\ \bibinfo {pages} {374} (\bibinfo {year}
  {2017})}\BibitemShut {NoStop}%
\bibitem [{\citenamefont {Yao}\ \emph {et~al.}(2021)\citenamefont {Yao},
  \citenamefont {Lin}, \citenamefont {Zhou},\ and\ \citenamefont
  {Lu}}]{yao_recent_2021}%
  \BibitemOpen
  \bibfield  {author} {\bibinfo {author} {\bibfnamefont {Y.}~\bibnamefont
  {Yao}}, \bibinfo {author} {\bibfnamefont {Q.}~\bibnamefont {Lin}}, \bibinfo
  {author} {\bibfnamefont {X.}~\bibnamefont {Zhou}},\ and\ \bibinfo {author}
  {\bibfnamefont {M.}~\bibnamefont {Lu}},\ }\bibfield  {title} {\bibinfo
  {title} {Recent research on the synthesis pentazolate anion
  \textit{cyclo}-{N}$_5^-$},\ }\href@noop {} {\bibfield  {journal} {\bibinfo
  {journal} {FirePhysChem}\ }\textbf {\bibinfo {volume} {1}},\ \bibinfo {pages}
  {33} (\bibinfo {year} {2021})}\BibitemShut {NoStop}%
\bibitem [{\citenamefont {Gelius}\ \emph {et~al.}(1971)\citenamefont {Gelius},
  \citenamefont {Allan}, \citenamefont {Johansson}, \citenamefont {Siegbahn},
  \citenamefont {Allison},\ and\ \citenamefont {Siegbahn}}]{gelius_esca_1971}%
  \BibitemOpen
  \bibfield  {author} {\bibinfo {author} {\bibfnamefont {U.}~\bibnamefont
  {Gelius}}, \bibinfo {author} {\bibfnamefont {C.~J.}\ \bibnamefont {Allan}},
  \bibinfo {author} {\bibfnamefont {G.}~\bibnamefont {Johansson}}, \bibinfo
  {author} {\bibfnamefont {H.}~\bibnamefont {Siegbahn}}, \bibinfo {author}
  {\bibfnamefont {D.~A.}\ \bibnamefont {Allison}},\ and\ \bibinfo {author}
  {\bibfnamefont {K.}~\bibnamefont {Siegbahn}},\ }\bibfield  {title} {\bibinfo
  {title} {The {ESCA} {Spectra} of {Benzene} and the {Iso}-electronic {Series},
  {Thiophene}, {Pyrrole} and {Furan}},\ }\href@noop {} {\bibfield  {journal}
  {\bibinfo  {journal} {Phys. Scr.}\ }\textbf {\bibinfo {volume} {3}},\
  \bibinfo {pages} {237} (\bibinfo {year} {1971})}\BibitemShut {NoStop}%
\bibitem [{\citenamefont {Pan}\ \emph {et~al.}(1986)\citenamefont {Pan},
  \citenamefont {Stair},\ and\ \citenamefont
  {Fleisch}}]{pan_chemisorption_1986}%
  \BibitemOpen
  \bibfield  {author} {\bibinfo {author} {\bibfnamefont {F.-M.}\ \bibnamefont
  {Pan}}, \bibinfo {author} {\bibfnamefont {P.~C.}\ \bibnamefont {Stair}},\
  and\ \bibinfo {author} {\bibfnamefont {T.~H.}\ \bibnamefont {Fleisch}},\
  }\bibfield  {title} {\bibinfo {title} {Chemisorption of pyridine and pyrrole
  on iron oxide surfaces studied by {XPS}},\ }\href@noop {} {\bibfield
  {journal} {\bibinfo  {journal} {Surface Science}\ }\textbf {\bibinfo {volume}
  {177}},\ \bibinfo {pages} {1} (\bibinfo {year} {1986})}\BibitemShut {NoStop}%
\bibitem [{\citenamefont {Cavell}\ and\ \citenamefont
  {Allison}(1977)}]{cavell1977site}%
  \BibitemOpen
  \bibfield  {author} {\bibinfo {author} {\bibfnamefont {R.~G.}\ \bibnamefont
  {Cavell}}\ and\ \bibinfo {author} {\bibfnamefont {D.~A.}\ \bibnamefont
  {Allison}},\ }\bibfield  {title} {\bibinfo {title} {Site of protonation in
  aromatic and acyclic amines and acyclic amides revealed by n1s core level
  electron spectroscopy},\ }\href@noop {} {\bibfield  {journal} {\bibinfo
  {journal} {Journal of the American Chemical Society}\ }\textbf {\bibinfo
  {volume} {99}},\ \bibinfo {pages} {4203} (\bibinfo {year}
  {1977})}\BibitemShut {NoStop}%
\bibitem [{\citenamefont {Nolting}\ \emph {et~al.}(2008)\citenamefont
  {Nolting}, \citenamefont {Ottosson}, \citenamefont {Faubel}, \citenamefont
  {Hertel},\ and\ \citenamefont {Winter}}]{nolting_pseudoequivalent_2008}%
  \BibitemOpen
  \bibfield  {author} {\bibinfo {author} {\bibfnamefont {D.}~\bibnamefont
  {Nolting}}, \bibinfo {author} {\bibfnamefont {N.}~\bibnamefont {Ottosson}},
  \bibinfo {author} {\bibfnamefont {M.}~\bibnamefont {Faubel}}, \bibinfo
  {author} {\bibfnamefont {I.~V.}\ \bibnamefont {Hertel}},\ and\ \bibinfo
  {author} {\bibfnamefont {B.}~\bibnamefont {Winter}},\ }\bibfield  {title}
  {\bibinfo {title} {Pseudoequivalent {Nitrogen} {Atoms} in {Aqueous}
  {Imidazole} {Distinguished} by {Chemical} {Shifts} in {Photoelectron}
  {Spectroscopy}},\ }\href@noop {} {\bibfield  {journal} {\bibinfo  {journal}
  {J. Am. Chem. Soc.}\ }\textbf {\bibinfo {volume} {130}},\ \bibinfo {pages}
  {8150} (\bibinfo {year} {2008})}\BibitemShut {NoStop}%
\bibitem [{\citenamefont {Schmidt}\ \emph {et~al.}(1993)\citenamefont
  {Schmidt}, \citenamefont {Baldridge}, \citenamefont {Boatz}, \citenamefont
  {Elbert}, \citenamefont {Gordon}, \citenamefont {Jensen}, \citenamefont
  {Koseki}, \citenamefont {Matsunaga}, \citenamefont {Nguyen}, \citenamefont
  {Su}, \citenamefont {Windus}, \citenamefont {Dupuis},\ and\ \citenamefont
  {Montgomery}}]{schmidt_general_1993}%
  \BibitemOpen
  \bibfield  {author} {\bibinfo {author} {\bibfnamefont {M.~W.}\ \bibnamefont
  {Schmidt}}, \bibinfo {author} {\bibfnamefont {K.~K.}\ \bibnamefont
  {Baldridge}}, \bibinfo {author} {\bibfnamefont {J.~A.}\ \bibnamefont
  {Boatz}}, \bibinfo {author} {\bibfnamefont {S.~T.}\ \bibnamefont {Elbert}},
  \bibinfo {author} {\bibfnamefont {M.~S.}\ \bibnamefont {Gordon}}, \bibinfo
  {author} {\bibfnamefont {J.~H.}\ \bibnamefont {Jensen}}, \bibinfo {author}
  {\bibfnamefont {S.}~\bibnamefont {Koseki}}, \bibinfo {author} {\bibfnamefont
  {N.}~\bibnamefont {Matsunaga}}, \bibinfo {author} {\bibfnamefont {K.~A.}\
  \bibnamefont {Nguyen}}, \bibinfo {author} {\bibfnamefont {S.}~\bibnamefont
  {Su}}, \bibinfo {author} {\bibfnamefont {T.~L.}\ \bibnamefont {Windus}},
  \bibinfo {author} {\bibfnamefont {M.}~\bibnamefont {Dupuis}},\ and\ \bibinfo
  {author} {\bibfnamefont {J.~A.}\ \bibnamefont {Montgomery}},\ }\bibfield
  {title} {\bibinfo {title} {General atomic and molecular electronic structure
  system},\ }\href@noop {} {\bibfield  {journal} {\bibinfo  {journal} {J.
  Comput. Chem.}\ }\textbf {\bibinfo {volume} {14}},\ \bibinfo {pages} {1347}
  (\bibinfo {year} {1993})}\BibitemShut {NoStop}%
\bibitem [{\citenamefont {Gordon}\ and\ \citenamefont
  {Schmidt}(2005)}]{gordon_advances_2005}%
  \BibitemOpen
  \bibfield  {author} {\bibinfo {author} {\bibfnamefont {M.~S.}\ \bibnamefont
  {Gordon}}\ and\ \bibinfo {author} {\bibfnamefont {M.~W.}\ \bibnamefont
  {Schmidt}},\ }\bibfield  {title} {\bibinfo {title} {Advances in electronic
  structure theory: {GAMESS} a decade later},\ }in\ \href@noop {} {\emph
  {\bibinfo {booktitle} {Theory and applications of computational chemistry}}}\
  (\bibinfo  {publisher} {Elsevier},\ \bibinfo {year} {2005})\ pp.\ \bibinfo
  {pages} {1167--1189}\BibitemShut {NoStop}%
\bibitem [{\citenamefont {Becke}(1988)}]{becke_density-functional_1988}%
  \BibitemOpen
  \bibfield  {author} {\bibinfo {author} {\bibfnamefont {A.~D.}\ \bibnamefont
  {Becke}},\ }\bibfield  {title} {\bibinfo {title} {Density-functional
  exchange-energy approximation with correct asymptotic behavior},\ }\href@noop
  {} {\bibfield  {journal} {\bibinfo  {journal} {Phys. Rev. A}\ }\textbf
  {\bibinfo {volume} {38}},\ \bibinfo {pages} {3098} (\bibinfo {year}
  {1988})}\BibitemShut {NoStop}%
\bibitem [{\citenamefont {Becke}(1993)}]{becke_new_1993}%
  \BibitemOpen
  \bibfield  {author} {\bibinfo {author} {\bibfnamefont {A.~D.}\ \bibnamefont
  {Becke}},\ }\bibfield  {title} {\bibinfo {title} {A new mixing of
  {Hartree}{\textendash}{Fock} and local density-functional theories},\
  }\href@noop {} {\bibfield  {journal} {\bibinfo  {journal} {J. Chem. Phys.}\
  }\textbf {\bibinfo {volume} {98}},\ \bibinfo {pages} {1372} (\bibinfo {year}
  {1993})}\BibitemShut {NoStop}%
\bibitem [{\citenamefont {Lee}\ \emph {et~al.}(1988)\citenamefont {Lee},
  \citenamefont {Yang},\ and\ \citenamefont {Parr}}]{lee_development_1988}%
  \BibitemOpen
  \bibfield  {author} {\bibinfo {author} {\bibfnamefont {C.}~\bibnamefont
  {Lee}}, \bibinfo {author} {\bibfnamefont {W.}~\bibnamefont {Yang}},\ and\
  \bibinfo {author} {\bibfnamefont {R.~G.}\ \bibnamefont {Parr}},\ }\bibfield
  {title} {\bibinfo {title} {Development of the {Colle}-{Salvetti}
  correlation-energy formula into a functional of the electron density},\
  }\href@noop {} {\bibfield  {journal} {\bibinfo  {journal} {Phys. Rev. B}\
  }\textbf {\bibinfo {volume} {37}},\ \bibinfo {pages} {785} (\bibinfo {year}
  {1988})}\BibitemShut {NoStop}%
\bibitem [{\citenamefont {Tian}\ \emph {et~al.}()\citenamefont {Tian},
  \citenamefont {Duan}, \citenamefont {Hua},\ and\ \citenamefont
  {Luo}}]{DynaVib}%
  \BibitemOpen
  \bibfield  {author} {\bibinfo {author} {\bibfnamefont {G.}~\bibnamefont
  {Tian}}, \bibinfo {author} {\bibfnamefont {S.}~\bibnamefont {Duan}}, \bibinfo
  {author} {\bibfnamefont {W.}~\bibnamefont {Hua}},\ and\ \bibinfo {author}
  {\bibfnamefont {Y.}~\bibnamefont {Luo}},\ }\href@noop {} {\bibinfo {title}
  {\uppercase{D}yna\uppercase{V}ib, version 1.0}},\ \bibinfo {note}
  {\uppercase{R}oyal Institute of Technology: Sweden, 2012}\BibitemShut
  {NoStop}%
\bibitem [{\citenamefont {Hua}\ \emph {et~al.}(2020)\citenamefont {Hua},
  \citenamefont {Tian},\ and\ \citenamefont {Luo}}]{hua_theoretical_2020}%
  \BibitemOpen
  \bibfield  {author} {\bibinfo {author} {\bibfnamefont {W.}~\bibnamefont
  {Hua}}, \bibinfo {author} {\bibfnamefont {G.}~\bibnamefont {Tian}},\ and\
  \bibinfo {author} {\bibfnamefont {Y.}~\bibnamefont {Luo}},\ }\bibfield
  {title} {\bibinfo {title} {Theoretical assessment of vibrationally resolved
  {C1s} {X}-ray photoelectron spectra of simple cyclic molecules},\ }\href@noop
  {} {\bibfield  {journal} {\bibinfo  {journal} {Phys. Chem. Chem. Phys.}\
  }\textbf {\bibinfo {volume} {22}},\ \bibinfo {pages} {20014} (\bibinfo {year}
  {2020})}\BibitemShut {NoStop}%
\bibitem [{\citenamefont {Duschinsky}(1937)}]{duschinsky_1937}%
  \BibitemOpen
  \bibfield  {author} {\bibinfo {author} {\bibfnamefont {F.}~\bibnamefont
  {Duschinsky}},\ }\href@noop {} {\bibfield  {journal} {\bibinfo  {journal}
  {Acta Physicochim. URSS}\ }\textbf {\bibinfo {volume} {7}},\ \bibinfo {pages}
  {551} (\bibinfo {year} {1937})}\BibitemShut {NoStop}%
\bibitem [{\citenamefont {Sharp}\ and\ \citenamefont
  {Rosenstock}(1964)}]{sharp_franckcondon_1964}%
  \BibitemOpen
  \bibfield  {author} {\bibinfo {author} {\bibfnamefont {T.}~\bibnamefont
  {Sharp}}\ and\ \bibinfo {author} {\bibfnamefont {H.}~\bibnamefont
  {Rosenstock}},\ }\bibfield  {title} {\bibinfo {title} {Franck-condon
  {Factors} for {Polyatomic} {Molecules}},\ }\href@noop {} {\bibfield
  {journal} {\bibinfo  {journal} {J. Chem. Phys.}\ }\textbf {\bibinfo {volume}
  {41}},\ \bibinfo {pages} {3453} (\bibinfo {year} {1964})}\BibitemShut
  {NoStop}%
\bibitem [{\citenamefont {Ruhoff}(1994)}]{ruhoff_recursion_1994}%
  \BibitemOpen
  \bibfield  {author} {\bibinfo {author} {\bibfnamefont {P.~T.}\ \bibnamefont
  {Ruhoff}},\ }\bibfield  {title} {\bibinfo {title} {Recursion relations for
  multi-dimensional {Franck}-{Condon} overlap integrals},\ }\href@noop {}
  {\bibfield  {journal} {\bibinfo  {journal} {Chem. Phys.}\ }\textbf {\bibinfo
  {volume} {186}},\ \bibinfo {pages} {355} (\bibinfo {year}
  {1994})}\BibitemShut {NoStop}%
\bibitem [{\citenamefont {Ruhoff}\ and\ \citenamefont
  {Ratner}(2000)}]{ruhoff_algorithms_2000}%
  \BibitemOpen
  \bibfield  {author} {\bibinfo {author} {\bibfnamefont {P.~T.}\ \bibnamefont
  {Ruhoff}}\ and\ \bibinfo {author} {\bibfnamefont {M.~A.}\ \bibnamefont
  {Ratner}},\ }\bibfield  {title} {\bibinfo {title} {Algorithms for computing
  {Franck}-{Condon} overlap integrals},\ }\href@noop {} {\bibfield  {journal}
  {\bibinfo  {journal} {Int. J. Quantum Chem.}\ }\textbf {\bibinfo {volume}
  {77}},\ \bibinfo {pages} {383} (\bibinfo {year} {2000})}\BibitemShut
  {NoStop}%
\bibitem [{\citenamefont {Triguero}\ \emph {et~al.}(1999)\citenamefont
  {Triguero}, \citenamefont {Plashkevych}, \citenamefont {Pettersson},\ and\
  \citenamefont {{\AA}gren}}]{triguero_separate_1999}%
  \BibitemOpen
  \bibfield  {author} {\bibinfo {author} {\bibfnamefont {L.}~\bibnamefont
  {Triguero}}, \bibinfo {author} {\bibfnamefont {O.}~\bibnamefont
  {Plashkevych}}, \bibinfo {author} {\bibfnamefont {L.}~\bibnamefont
  {Pettersson}},\ and\ \bibinfo {author} {\bibfnamefont {H.}~\bibnamefont
  {{\AA}gren}},\ }\bibfield  {title} {\bibinfo {title} {Separate state vs.
  transition state {Kohn}-{Sham} calculations of {X}-ray photoelectron binding
  energies and chemical shifts},\ }\href@noop {} {\bibfield  {journal}
  {\bibinfo  {journal} {J. Electron Spectrosc.}\ }\textbf {\bibinfo {volume}
  {104}},\ \bibinfo {pages} {195} (\bibinfo {year} {1999})}\BibitemShut
  {NoStop}%
\bibitem [{\citenamefont {Bagus}\ \emph {et~al.}(2016)\citenamefont {Bagus},
  \citenamefont {Sousa},\ and\ \citenamefont
  {Illas}}]{bagus_consequences_2016}%
  \BibitemOpen
  \bibfield  {author} {\bibinfo {author} {\bibfnamefont {P.~S.}\ \bibnamefont
  {Bagus}}, \bibinfo {author} {\bibfnamefont {C.}~\bibnamefont {Sousa}},\ and\
  \bibinfo {author} {\bibfnamefont {F.}~\bibnamefont {Illas}},\ }\bibfield
  {title} {\bibinfo {title} {Consequences of electron correlation for {XPS}
  binding energies: {Representative} case for {C}(1s) and {O}(1s) {XPS} of
  {CO}},\ }\href@noop {} {\bibfield  {journal} {\bibinfo  {journal} {J. Chem.
  Phys.}\ }\textbf {\bibinfo {volume} {145}},\ \bibinfo {pages} {144303}
  (\bibinfo {year} {2016})}\BibitemShut {NoStop}%
\bibitem [{\citenamefont {Bellafont}\ \emph
  {et~al.}(2015{\natexlab{a}})\citenamefont {Bellafont}, \citenamefont
  {Illas},\ and\ \citenamefont {Bagus}}]{pueyo_bellafont_validation_2015}%
  \BibitemOpen
  \bibfield  {author} {\bibinfo {author} {\bibfnamefont {N.~P.}\ \bibnamefont
  {Bellafont}}, \bibinfo {author} {\bibfnamefont {F.}~\bibnamefont {Illas}},\
  and\ \bibinfo {author} {\bibfnamefont {P.~S.}\ \bibnamefont {Bagus}},\
  }\bibfield  {title} {\bibinfo {title} {Validation of {Koopmans}' theorem for
  density functional theory binding energies},\ }\href@noop {} {\bibfield
  {journal} {\bibinfo  {journal} {Phys. Chem. Chem. Phys.}\ }\textbf {\bibinfo
  {volume} {17}},\ \bibinfo {pages} {4015} (\bibinfo {year}
  {2015}{\natexlab{a}})}\BibitemShut {NoStop}%
\bibitem [{\citenamefont {Bellafont}\ \emph
  {et~al.}(2015{\natexlab{b}})\citenamefont {Bellafont}, \citenamefont
  {Bagus},\ and\ \citenamefont {Illas}}]{pueyo_bellafont_prediction_2015}%
  \BibitemOpen
  \bibfield  {author} {\bibinfo {author} {\bibfnamefont {N.~P.}\ \bibnamefont
  {Bellafont}}, \bibinfo {author} {\bibfnamefont {P.~S.}\ \bibnamefont
  {Bagus}},\ and\ \bibinfo {author} {\bibfnamefont {F.}~\bibnamefont {Illas}},\
  }\bibfield  {title} {\bibinfo {title} {Prediction of core level binding
  energies in density functional theory: {Rigorous} definition of initial and
  final state contributions and implications on the physical meaning of
  {Kohn}-{Sham} energies},\ }\href@noop {} {\bibfield  {journal} {\bibinfo
  {journal} {J. Chem. Phys.}\ }\textbf {\bibinfo {volume} {142}},\ \bibinfo
  {pages} {214102} (\bibinfo {year} {2015}{\natexlab{b}})}\BibitemShut
  {NoStop}%
\bibitem [{\citenamefont {Pueyo~Bellafont}\ \emph {et~al.}(2016)\citenamefont
  {Pueyo~Bellafont}, \citenamefont {Vi{\~n}es},\ and\ \citenamefont
  {Illas}}]{pueyo_SCF_performance_2016}%
  \BibitemOpen
  \bibfield  {author} {\bibinfo {author} {\bibfnamefont {N.}~\bibnamefont
  {Pueyo~Bellafont}}, \bibinfo {author} {\bibfnamefont {F.}~\bibnamefont
  {Vi{\~n}es}},\ and\ \bibinfo {author} {\bibfnamefont {F.}~\bibnamefont
  {Illas}},\ }\bibfield  {title} {\bibinfo {title} {Performance of the {TPSS}
  {Functional} on {Predicting} {Core} {Level} {Binding} {Energies} of {Main}
  {Group} {Elements} {Containing} {Molecules}: {A} {Good} {Choice} for
  {Molecules} {Adsorbed} on {Metal} {Surfaces}},\ }\href@noop {} {\bibfield
  {journal} {\bibinfo  {journal} {J. Chem. Theory Comput.}\ }\textbf {\bibinfo
  {volume} {12}},\ \bibinfo {pages} {324} (\bibinfo {year} {2016})}\BibitemShut
  {NoStop}%
\bibitem [{\citenamefont {Du}\ \emph {et~al.}(2022)\citenamefont {Du},
  \citenamefont {Wang}, \citenamefont {Wei}, \citenamefont {Zhang},
  \citenamefont {Ge},\ and\ \citenamefont {Hua}}]{du_theoretical_2022}%
  \BibitemOpen
  \bibfield  {author} {\bibinfo {author} {\bibfnamefont {X.}~\bibnamefont
  {Du}}, \bibinfo {author} {\bibfnamefont {S.-Y.}\ \bibnamefont {Wang}},
  \bibinfo {author} {\bibfnamefont {M.}~\bibnamefont {Wei}}, \bibinfo {author}
  {\bibfnamefont {J.-R.}\ \bibnamefont {Zhang}}, \bibinfo {author}
  {\bibfnamefont {G.}~\bibnamefont {Ge}},\ and\ \bibinfo {author}
  {\bibfnamefont {W.}~\bibnamefont {Hua}},\ }\bibfield  {title} {\bibinfo
  {title} {A theoretical library of {N1s} core binding energies of polynitrogen
  molecules and ions in the gas phase},\ }\href@noop {} {\bibfield  {journal}
  {\bibinfo  {journal} {Phys. Chem. Chem. Phys.}\ }\textbf {\bibinfo {volume}
  {24}},\ \bibinfo {pages} {8196} (\bibinfo {year} {2022})}\BibitemShut
  {NoStop}%
\bibitem [{\citenamefont
  {Atahan-Evrenk}(2018)}]{atahan-evrenk_quantitative_2018}%
  \BibitemOpen
  \bibfield  {author} {\bibinfo {author} {\bibfnamefont {S.}~\bibnamefont
  {Atahan-Evrenk}},\ }\bibfield  {title} {\bibinfo {title} {A quantitative
  structure{\textendash}property study of reorganization energy for known
  p-type organic semiconductors},\ }\href@noop {} {\bibfield  {journal}
  {\bibinfo  {journal} {RSC Adv.}\ }\textbf {\bibinfo {volume} {8}},\ \bibinfo
  {pages} {40330} (\bibinfo {year} {2018})}\BibitemShut {NoStop}%
\bibitem [{si_()}]{si_five}%
  \BibitemOpen
  \href@noop {} {}\bibinfo {note} {See Supplemental Material at[URL will be
  inserted by publisher] for visualization of amine and imine IPs, theoretical
  spectra of 12 individual five-membered N-heterocycles with analysis;
  additional examples for effects of hydrogenation and benzene rings,
  additional vibrational analyses, and energies of all 35
  N-heterocycles.}\BibitemShut {Stop}%
\bibitem [{\citenamefont {Qiu}\ \emph {et~al.}(2021)\citenamefont {Qiu},
  \citenamefont {Tian}, \citenamefont {Lin}, \citenamefont {Pan}, \citenamefont
  {Ye}, \citenamefont {Wang}, \citenamefont {Ma}, \citenamefont {Hu},
  \citenamefont {Luo},\ and\ \citenamefont {Ma}}]{qiu_narrowband_2021}%
  \BibitemOpen
  \bibfield  {author} {\bibinfo {author} {\bibfnamefont {X.}~\bibnamefont
  {Qiu}}, \bibinfo {author} {\bibfnamefont {G.}~\bibnamefont {Tian}}, \bibinfo
  {author} {\bibfnamefont {C.}~\bibnamefont {Lin}}, \bibinfo {author}
  {\bibfnamefont {Y.}~\bibnamefont {Pan}}, \bibinfo {author} {\bibfnamefont
  {X.}~\bibnamefont {Ye}}, \bibinfo {author} {\bibfnamefont {B.}~\bibnamefont
  {Wang}}, \bibinfo {author} {\bibfnamefont {D.}~\bibnamefont {Ma}}, \bibinfo
  {author} {\bibfnamefont {D.}~\bibnamefont {Hu}}, \bibinfo {author}
  {\bibfnamefont {Y.}~\bibnamefont {Luo}},\ and\ \bibinfo {author}
  {\bibfnamefont {Y.}~\bibnamefont {Ma}},\ }\bibfield  {title} {\bibinfo
  {title} {Narrowband {Emission} from {Organic} {Fluorescent} {Emitters} with
  {Dominant} {Low}-{Frequency} {Vibronic} {Coupling}},\ }\href@noop {}
  {\bibfield  {journal} {\bibinfo  {journal} {Advanced Optical Materials}\
  }\textbf {\bibinfo {volume} {9}},\ \bibinfo {pages} {2001845} (\bibinfo
  {year} {2021})}\BibitemShut {NoStop}%
\bibitem [{\citenamefont {Greczynski}\ and\ \citenamefont
  {Hultman}(2020)}]{greczynski_x-ray_2020}%
  \BibitemOpen
  \bibfield  {author} {\bibinfo {author} {\bibfnamefont {G.}~\bibnamefont
  {Greczynski}}\ and\ \bibinfo {author} {\bibfnamefont {L.}~\bibnamefont
  {Hultman}},\ }\bibfield  {title} {\bibinfo {title} {X-ray photoelectron
  spectroscopy: {Towards} reliable binding energy referencing},\ }\href@noop {}
  {\bibfield  {journal} {\bibinfo  {journal} {Prog. Mater. Sci.}\ }\textbf
  {\bibinfo {volume} {107}},\ \bibinfo {pages} {100591} (\bibinfo {year}
  {2020})}\BibitemShut {NoStop}%
\bibitem [{\citenamefont {Nicolas}\ and\ \citenamefont
  {Miron}(2012)}]{nicolas_lifetime_2012}%
  \BibitemOpen
  \bibfield  {author} {\bibinfo {author} {\bibfnamefont {C.}~\bibnamefont
  {Nicolas}}\ and\ \bibinfo {author} {\bibfnamefont {C.}~\bibnamefont
  {Miron}},\ }\bibfield  {title} {\bibinfo {title} {Lifetime broadening of
  core-excited and -ionized states},\ }\href@noop {} {\bibfield  {journal}
  {\bibinfo  {journal} {J. Electron. Spectrosc. Relat. Phenom.}\ }\textbf
  {\bibinfo {volume} {185}},\ \bibinfo {pages} {267} (\bibinfo {year}
  {2012})}\BibitemShut {NoStop}%
\bibitem [{\citenamefont {Chen}\ \emph {et~al.}(1989)\citenamefont {Chen},
  \citenamefont {Ma},\ and\ \citenamefont {Sette}}]{chen_k_1989}%
  \BibitemOpen
  \bibfield  {author} {\bibinfo {author} {\bibfnamefont {C.~T.}\ \bibnamefont
  {Chen}}, \bibinfo {author} {\bibfnamefont {Y.}~\bibnamefont {Ma}},\ and\
  \bibinfo {author} {\bibfnamefont {F.}~\bibnamefont {Sette}},\ }\bibfield
  {title} {\bibinfo {title} {\textit{{K}} -shell photoabsorption of the {N}$_2$
  molecule},\ }\href@noop {} {\bibfield  {journal} {\bibinfo  {journal} {Phys.
  Rev. A}\ }\textbf {\bibinfo {volume} {40}},\ \bibinfo {pages} {6737}
  (\bibinfo {year} {1989})}\BibitemShut {NoStop}%
\bibitem [{\citenamefont {Giertz}\ \emph {et~al.}(2002)\citenamefont {Giertz},
  \citenamefont {B{\o}rve}, \citenamefont {B{\"a}{\ss}ler}, \citenamefont
  {Wiesner}, \citenamefont {Svensson}, \citenamefont {Karlsson},\ and\
  \citenamefont {S{\ae}thre}}]{giertz_vibrationally_2002}%
  \BibitemOpen
  \bibfield  {author} {\bibinfo {author} {\bibfnamefont {A.}~\bibnamefont
  {Giertz}}, \bibinfo {author} {\bibfnamefont {K.}~\bibnamefont {B{\o}rve}},
  \bibinfo {author} {\bibfnamefont {M.}~\bibnamefont {B{\"a}{\ss}ler}},
  \bibinfo {author} {\bibfnamefont {K.}~\bibnamefont {Wiesner}}, \bibinfo
  {author} {\bibfnamefont {S.}~\bibnamefont {Svensson}}, \bibinfo {author}
  {\bibfnamefont {L.}~\bibnamefont {Karlsson}},\ and\ \bibinfo {author}
  {\bibfnamefont {L.}~\bibnamefont {S{\ae}thre}},\ }\bibfield  {title}
  {\bibinfo {title} {Vibrationally resolved photoelectron spectra of the carbon
  1s and nitrogen 1s shells in hydrogen cyanide},\ }\href@noop {} {\bibfield
  {journal} {\bibinfo  {journal} {Chem. Phys.}\ }\textbf {\bibinfo {volume}
  {277}},\ \bibinfo {pages} {83} (\bibinfo {year} {2002})}\BibitemShut
  {NoStop}%
\bibitem [{\citenamefont {Katrib}\ \emph {et~al.}(1983)\citenamefont {Katrib},
  \citenamefont {El-Rayyes},\ and\ \citenamefont {Al-Kharafi}}]{katrib_n_1983}%
  \BibitemOpen
  \bibfield  {author} {\bibinfo {author} {\bibfnamefont {A.}~\bibnamefont
  {Katrib}}, \bibinfo {author} {\bibfnamefont {N.}~\bibnamefont {El-Rayyes}},\
  and\ \bibinfo {author} {\bibfnamefont {F.}~\bibnamefont {Al-Kharafi}},\
  }\bibfield  {title} {\bibinfo {title} {N 1s orbital binding energies of some
  pyrazole pyrazoline compounds by {XPS}},\ }\href@noop {} {\bibfield
  {journal} {\bibinfo  {journal} {J. Electron. Spectrosc. Relat. Phenom.}\
  }\textbf {\bibinfo {volume} {31}},\ \bibinfo {pages} {317} (\bibinfo {year}
  {1983})}\BibitemShut {NoStop}%
\end{thebibliography}

%========================TABLES ====================================

\begin{table*}
    \centering
        \caption{
Vertical and adiabatic ionization potentials ($I^\text{vert}$ and $I^\text{ad}$), their difference ($\Delta I$), the 0-0 transition energy ($E^\text{00}_\text{DR}$), and the ZPE difference ($\bigtriangleup {\varepsilon}_{0}$) [see Eqs. (\ref{eq:Iv})--(\ref{eq:deps0})] of all systems simulated with B3LYP.  All energies are in eV.  N$^*$ denotes the core-ionized N atom; amine (bold) and imine (normal) N atoms are indicated by different fonts. Relative deviations to experiments are given in parentheses. }   
    \resizebox{\textwidth}{!}{
    \begin{ruledtabular}
\begin{threeparttable}
    \begin{tabular}{rccccccccc}
        Molecule & N$^*$ & Expt.  & $I^\text{vert}$ & $I^\text{ad}$ & $\Delta I$ 
 &$E^\text{DR}_\text{00}$ & $\bigtriangleup {\varepsilon}_{0}$   \\ \hline
        2H-pyrrole & N & ~ & 404.20 & 404.06 & 0.14& 404.10 & 0.04  \\ 
       1-Pyrroline & N & ~ & 404.37 & 404.21 &0.16&  404.23 & 0.02  \\     
       3-Pyrroline & \textbf{N} & ~ & 404.50 & 404.21& 0.29 & 404.23 & 0.02  \\ 
       2-Pyrroline & \textbf{N}& ~ & 404.84 & 404.57 &0.27& 404.57 & 0.00  \\ 
        1H-pyrrole & \textbf{N} & 406.15\tnote{a} /406.10\tnote{b} & 406.03(-0.12/0.07) & 405.76(-0.39/-0.34) & 0.27& 405.73 & -0.03  \\ 
        &  &406.00\tnote{c} /406.18\tnote{d}& +0.03/-0.15)&-0.24/-0.42)& &    \\
        Imidazole & \textbf{N1} & ~405.60\tnote{e} & 406.59 (+0.99) & 406.23 (+0.63) & 0.36& 406.16 & -0.07  \\ 
    ~ & N2 & 403.90\tnote{e} & 404.25 (+0.35) & 404.06 (+0.16) &0.19& 404.10 & 0.04 \\ 
       Pyrazole & \textbf{N1} & ~ & 406.83 & 406.45 &0.38& 406.38 & -0.07  \\ 
        ~ & N2 & ~ & 405.10 & 404.94 &0.16& 404.96 & 0.02  \\ 
        2-Pyrazoline & \textbf{N1} & ~ & 405.41 & 405.06 &0.35& 405.03 & -0.03  \\ 
        ~ & N2 & ~ & 405.32 & 405.16 &0.16& 405.16 & 0.00  \\ 
        1, 3, 4-Triazole & \textbf{N1} & ~ & 407.23 & 406.86 &0.37& 406.78 & -0.08  \\ 
        ~ & N2 & ~ & 405.21 & 405.04 & 0.17& 405.06 & 0.02  \\ 
        1, 2, 5-Triazole & \textbf{N1} & ~ & 407.87 & 407.39 &0.48& 407.3 & -0.09  \\ 
        ~ & N2 & ~ & 405.98 & 405.80 & 0.18&405.79 & -0.01  \\ 
        1, 2, 4-Triazole & \textbf{N1} & ~ & 407.44 & 407.02 &0.42& 406.93 & -0.09  \\ 
        ~ & N2 & ~ & 405.71 & 405.53 & 0.18& 405.53 & 0.00  \\ 
        ~ & N3 & ~ & 404.91 & 404.69 &0.22& 404.71 & 0.02  \\ 
         \textit{Cyclo}-N$_5^-$ &N &  & 400.58 & 400.28 &0.30& 399.98 & -0.03  \\ 
    \end{tabular}
    \begin{tablenotes}
\item[a] Jolly et al.\cite{jolly_core-electron_1984}
\item[b] Gelius et al.\cite{gelius_esca_1971}
\item[c] Pan et al.\cite{pan_chemisorption_1986}
\item[d] Cavell et al. \cite{cavell1977site}
\item[e] Nolting et al. \cite{nolting_pseudoequivalent_2008}
         \end{tablenotes} 
         \end{threeparttable}
\end{ruledtabular}
    \label{tab:ip}
    }
\end{table*}

    %%%%%%%%%%%%%%%%%%%%%%%Table structure
\begin{table*}
    \centering
    \caption{Structural changes of all molecules in their optimized FCH state (\textbf{min FCH}) compared to the optimized ground state (\textbf{min GS}). Selected bond lengths (in {\AA}) and angles (in$^\circ$)  near the ionized nitrogen (N$^*$) are listed.  The RMSD (in \AA) of the two superimposed structures is also given (aligned with the smallest RMSD value).
    }    \label{tab:st}
\resizebox{\textwidth}{!}{
\begin{threeparttable}

    \begin{tabular}{rlcccccc}
    \hline\hline 
        Molecules & N$^*$  & N$_\text{a}^*$-C & N$_\text{a}^*$-H & N$_\text{i}^*$-C&$\angle$C-N$_\text{a}^*$-C & $\angle$C-N$_\text{i}^*$-C & RMSD  \\ \hline
2H-pyrrole & N           & ~           &            & 1.44(-0.02); 1.29(+0.00) & & 109.2(+3.2) & 0.03   \\     
1-Pyrroline & N          & ~           &            &  1.50(+0.03); 1.25(-0.01) &   & 112.3(+3.5) & 0.04   \\
3-Pyrroline & \textbf{N} & 1.50(+0.03); 1.50(+0.03) & 0.97(-0.04) &   & 112.9(+3.5) & & 0.10   \\ 
2-Pyrroline & \textbf{N} & 1.46(+0.06); 1.52(+0.04) & 0.97(-0.04)  & ~ & 109.0(+3.0) &  & 0.15   \\ 
1H-pyrrole  & \textbf{N} & 1.44(+0.07); 1.44(+0.07) &  0.98(-0.02) &  ~ &108.7(-1.2) &  & 0.03   \\ 
Imidazole & \textbf{N1}  &  1.49(+0.13); 1.43(+0.05) & 0.98(-0.02)  & ~  & 104.7(-2.6) & ~ & 0.05   \\ 
         ~ & N2          & ~           &           & 1.37(+0.00); 1.31(+0.00)  & ~ & 109.1(+3.6) & 0.04   \\ 
Pyrazole & \textbf{N1}   & 1.51(+0.17)\tnote{a}; 1.40(+0.04) &0.98(-0.03) &  & 110.5(-2.7)\tnote{c} & & 0.05   \\ 
      ~ & N2             &   & ~ & 1.34(+0.00)\tnote{b}; 1.34(+0.01)  & ~ &  107.0(+2.8)\tnote{e}  &  0.04   \\ 
2-Pyrazoline & \textbf{N1}& 1.55(+0.15)\tnote{a}; 1.48(+0.00) & 0.98(-0.02) &   &110.8(+2.3)\tnote{c}  & ~ &0.09   \\ 
           ~ & N2        &   &        & 1.45(+0.06)\tnote{b}; 1.26(-0.02) &   & 111.2(+2.4)\tnote{e} &  0.04   \\ 
1, 3, 4-Triazole & \textbf{N1} &1.46(+0.09); 1.46(+0.09) & 0.98(-0.02)&   & 102.1(-2.5) & ~ & 0.04   \\ 
        ~ & N2 &   & ~ & 1.41(+0.03)\tnote{b}; 1.29(-0.02)  &    &  110.3(+2.9)\tnote{e}  & 0.03   \\ 
1, 2, 5-Triazole & \textbf{N1} & 1.45(+0.12)\tnote{a}; 1.45(+0.12)\tnote{a}& 0.98(-0.02)& & 113.7(-2.9)\tnote{d}  & ~ & 0.07   \\ 
        ~ & N2 &   & ~ & 1.33(-0.00); 1.34(+0.00)\tnote{b} & ~  &  105.2(+2.5)\tnote{e}  & 0.04   \\ 
1, 2, 4-Triazole & \textbf{N1} & 1.50(+0.14)\tnote{a}; 1.42(+0.07) & 0.98(-0.02) & ~ & 107.2(-3.1)\tnote{c} & ~ & 0.05   \\ 
        ~ & N2 &   & ~ & 1.35(-0.01)\tnote{b}; 1.34(+0.02)  & ~ & 104.4(+2.5)\tnote{e} & 0.04   \\ 
        ~ & N3 & ~ & ~ &  1.37(+0.00); 1.31(+0.00) & ~ & 106.2(+3.4) & 0.04   \\ 
   %\textit{Cyclo}-N$_5^-$& N & 1.37(+0.05)\tnote{a}; 1.37(+0.05)\tnote{a}&~& &107.3(-0.7)&~& 0.05\\
   \hline\hline   \end{tabular}
        
\begin{tablenotes}
\item[a] N$_\text{a}^*$-N.
\item[b] N$_\text{i}^*$-N.
\item[c] $\angle$C-N$_\text{a}^*$-N.
\item[d] $\angle$N-N$_\text{a}^*$-N.
\item[e] $\angle$C-N$_\text{i}^*$-N.
\end{tablenotes}
\end{threeparttable}
}
\end{table*}

%%%%%%%%%%%%%%%%%%%%%%%Table DR
\begin{table}
    \centering
        \caption{Analysis of selected vibrational modes with large (with a threshold $S_\text{i} \geq$ 0.3) Huang-Rhys factors for the 11 five-membered molecules. Vibrational frequencies ${\omega}_i$  and the total vibrational reorganization energy ($E_\text{r}$) of the excited (FCH) state are given.  Amine (bold) and imine (lightface) N atoms are in different fonts.
     }   
    \begin{ruledtabular}
\begin{threeparttable}
    \begin{tabular}{lccccc}
Molecule &N$^*$& $i$& $\omega_i$ (cm$^{-1}$)& $S_\text{i}$& $E_\text{r}$ (eV)   \\     \hline
2H-pyrrole   &   N        & 4 & 849.1 & 0.70& 0.14      \\    \hline  
1-Pyrroline  &   N        & 5 & 804.5 & 0.41& 0.15      \\      \hline    
3-Pyrroline  &\textbf{N}  & 3 & 406.5 & 1.39& 0.59      \\         
             &            & 19&1401.8 & 0.85&           \\           
             &            & 30&3855.0 & 0.50&           \\       \hline  
2-Pyrroline  &\textbf{N}  & 1 & 80.6  &0.65 & 0.32      \\        
             &            & 6 &759.2  &0.33 &      \\              
             &            &20 &1425.7 &0.36 &      \\     \hline      
1H-pyrrole   &\textbf{N}  &12 &1036.9 &0.40 &0.34  \\
             &            & 18&1558.7 & 1.03&       \\ \hline              
Imidazole    &\textbf{N1}  & 9 &936.1  & 0.36&0.58  \\         
             &            & 10& 945.8 & 0.31&       \\            
             &            &12 &1155.4 & 0.58&       \\           
             &            &16 &1592.1 & 1.73&       \\       
             &    N2      & 7 &935.7  & 0.72&0.18    \\   \hline         
Pyrazole    &\textbf{N1}  & 8 &917.6  & 0.59&0.61   \\         
             &            & 9  & 969.9 &1.32&       \\           
             &            &  12& 1144.4&0.96&       \\        
             &            & 13 & 1226.3&0.36&       \\            
             &            & 14 & 1280.8&0.59&       \\          
             &   N2       &  8 & 976.6 &0.52&0.16   \\     \hline          
2-Pyrazoline&\textbf{N1}  & 1  & 144.0 &0.37&0.51 \\         
             &            & 6  & 800.7 &0.70&       \\           
             &            & 10 & 932.2 &1.04&       \\          
             &            & 17 &1339.2 &0.50&       \\ 
             &   N2       & 4  &662.1  &0.33&0.16   \\           
             &            & 6  &852.6  &0.46&       \\   \hline
1,3,4-Triazole&\textbf{N1}& 10 &945.9  &0.82&0.52   \\        
             &            & 15 &1625.7 &1.76&       \\        
             &    N2      &  6 & 954.1 &0.56&0.17\\ 
             &            &  9 &1082.2 &0.30&       \\  \hline    
1,2,5-Triazole&\textbf{N1}& 7  &935.2  &1.95&0.63  \\        
             &            & 8  & 948.1 &0.45&       \\       
             &            & 10 &1024.0 &1.90&       \\       
             &            & 12 &1273.1 &0.46&       \\
             &     N2     & 7  & 989.0 &0.70&0.18   \\   \hline     
1,2,4-Triazole&\textbf{N1}& 7  &901.6  &0.72&0.53   \\     
             &            &  8 &  923.9&1.27&       \\      
             &            & 10 & 1107.5&0.50&       \\       
             &            & 13 &1286.2 &0.39&       \\      
             &            & 14 & 1540.8&0.32&       \\ 
             &       N2   & 7  & 1008.5&0.59&0.18 \\    
             &      N3    & 6  & 961.5&0.84 &0.10  \\      
    \end{tabular}
         \end{threeparttable}
\end{ruledtabular}
    \label{si}
    \end{table}

%%%%%%%%%%%%%%%%%%%%%%%Table 4

\begin{table}
    \centering
 \caption{The total vibrational reorganization energy ($E_\text{r}$) and transition quantum number $n$ (when the spectra achieve convergence) for three groups of single and bicyclic molecules.
        }   
    \begin{ruledtabular}
\begin{threeparttable}
    \begin{tabular}{lccc}
Molecule &N$^*$& $E_\text{r}$ (eV) &  $n$     \\     \hline
Pyrrole   & N  & 0.34 & 6      \\   
Indole    & N  & 0.46 & 7    \\      \hline       
2H-pyrrole &N  &0.14 &5   \\
1H-isoindole&N &0.16 &6      \\ \hline              
Imidazole   &N1& 0.58 & 7  \\              
            & N2  & 0.18 &5  \\       
Benzimidazole &N1  & 0.86 &8    \\                  
             &   N2       &  0.21& 6   \\           
    \end{tabular}
%    \begin{tablenotes}
%         \end{tablenotes} 
         \end{threeparttable}
\end{ruledtabular}
    \label{erduibi}
    \end{table}

%%%%%%%%%%%%%%%%%figure%%%%%%%%%%%%%%%%%%

%%%%%%%%%%%%%%%%%figure%%%%%%%%%%%%%%%%%%

%%%%%%%%%%%%%%%%%figure%%%%%%%%%%%%%%%%%%

%aligned by (a) amine and (b) imine nitrogens, respectively.
%%%%%%%%%%%%%%%%%figure%%%%%%%%%%%%%%%%%%
%%%%%%%%%%%%%%%%%figure%%%%%%%%%%%%%%%%%%

%%%%%%%%%%%%%%%%%figure%%%%%%%%%%%%%%%%%%

%%%%%%%%%%%%%%%%%figure%%%%%%%%%%%%%%%%%%

%%%%%%%%%%%%%%%%%figure%%%%%%%%%%%%%%%%%%

%%%%%%%%%%%%%%%%%figure%%%%%%%%%%%%%%%%%%

\nocite{*}

%\end{CJK*}
\end{document}